\documentclass[preprint2]{aastex}

\usepackage{natbib}
\usepackage[breaklinks=true]{hyperref}
\usepackage{color}
\usepackage{epsfig}
\usepackage{textcomp}
\usepackage{amsmath}
\usepackage{graphicx}
\usepackage{epstopdf}

\shorttitle{Parameterizing the SN Ia Rate as a function of Host Galaxy Properties}
\shortauthors{M.~Smith et. al.}

\newcommand\PEGASE{P\'EGASE.2 }
\newcommand\mgal{\ifmmode M \else $M \,$\fi}
\newcommand\av{\ifmmode A_V \else $A_V \,$\fi}
\newcommand{\degree}{\ensuremath{^\circ}} 
\newcommand\vsurvey{\ifmmode V_{\textrm{survey}} \else $V_{\textrm{survey}}$\fi}
\newcommand\omatter{\ifmmode \Omega_{M}\else $\Omega_{M}$\fi}
\newcommand\olambda{\ifmmode \Omega_{\Lambda}\else $\Omega_{\Lambda}$\fi}
\newcommand\mperyr{\ifmmode \mathrm{M}_{\odot}\,\mathrm{yr}^{-1} \else $\mathrm{M}_{\odot}\,\mathrm{yr}^{-1}$ \fi}
\newcommand\vmax{\ifmmode V_{\mathrm{max}}\else $V_{\mathrm{max}}$\fi}
\newcommand\msun{\ifmmode \mathrm{M}_{\odot} \else $\mathrm{M}_{\odot}$\fi}
\newcommand\snrIa{\ifmmode \mathrm{SNR}_{\mathrm{Ia}} \else $\mathrm{SNR}_{\mathrm{Ia}}$\fi}
\newcommand\nmass{\ifmmode \mathrm{n}_{\mathrm{M}} \else $\mathrm{n}_{\mathrm{M}}$\fi}
\newcommand\nsfr{\ifmmode \mathrm{n}_{\mathrm{SFR}} \else $\mathrm{n}_{\mathrm{SFR}}$\fi}
\newcommand\Aunits{\ifmmode \mathrm{SNe}\,\,\mathrm{yr}^{-1}\,\msun^{-1} \else $\mathrm{SNe}\,\,\mathrm{yr}^{-1}\,\msun^{-1}$ \fi}
\newcommand\Bunits{\ifmmode \mathrm{SNe}\,\,\mathrm{yr}^{-1}\,(\mperyr)^{-1} \else $\mathrm{SNe}\,\,\mathrm{yr}^{-1}\,(\mperyr)^{-1}$ \fi}

\begin{document}

\title{The SDSS-II Supernova Survey: Parameterizing the Type Ia Supernova Rate as a Function of Host Galaxy Properties}

\author{Mathew~Smith\altaffilmark{1,2,3}
	 Robert~C~Nichol\altaffilmark{2}, 
	 Benjamin~Dilday\altaffilmark{4,5},
	 John~Marriner\altaffilmark{6},
	 Richard~Kessler\altaffilmark{7,8},
	 Bruce~Bassett\altaffilmark{3,9,10},
	 David~Cinabro\altaffilmark{11}, 
	 Joshua~Frieman\altaffilmark{6,7,8}, 
	 Peter~Garnavich\altaffilmark{12},
	 Saurabh~W~Jha\altaffilmark{13},
	 Hubert~Lampeitl\altaffilmark{2}, 
	 Masao~Sako \altaffilmark{14},
	 Donald~P~Schneider\altaffilmark{15},
	 Jesper~Sollerman\altaffilmark{16}}

\altaffiltext{1}{Astrophysics, Cosmology and Gravity Centre (ACGC), Department of Mathematics and Applied Mathematics, University of Cape Town, Rondebosch, 7701, SA}
\altaffiltext{2}{Institute of Cosmology and Gravitation, University of Portsmouth, Portsmouth, PO1 3FX, UK}
\altaffiltext{3}{African Institute for Mathematical Sciences, 6-8 Melrose Road, Muizenberg 7945, SA}
\altaffiltext{4}{Las Cumbres Observatory Global Telescope Network, 6740 Cortona Dr., Suite 102, Goleta, CA 93117, USA}
\altaffiltext{5}{Department of Physics, University of California, Santa Barbara, Broida Hall, Mail Code 9530, Santa Barbara, CA 93106-9530, USA}
\altaffiltext{6}{Center for Particle Astrophysics, Fermilab, P.O. Box 500, Batavia, IL 60510, USA}
\altaffiltext{7}{Department of Astronomy \& Astrophysics, The University of Chicago, 5640 S. Ellis Ave, Chicago, IL 60637, USA}
\altaffiltext{8}{Kavli Institute for Cosmological Physics, The University of Chicago, 5640 S. Ellis Ave, Chicago, IL 60637, USA}
\altaffiltext{9}{South African Astronomical Observatory, P.O. Box 9, Observatory 7935, South Africa}
\altaffiltext{10}{Department of Mathematics and Applied Mathematics, University of Cape Town, Rondebosch, 7701, SA}
\altaffiltext{11}{Wayne State University, Department of Physics and Astronomy, Detroit, MI, 48201, USA}
\altaffiltext{12}{Department of Physics, University of Notre Dame, 225 Nieuwland Science Hall, Notre Dame, IN 46556, USA}
\altaffiltext{13}{Department of Physics and Astronomy, Rutgers, the State University of New Jersey, 136 Frelinghuysen Road, Piscataway, NJ 08854, USA}
\altaffiltext{14}{Physics \& Astronomy, University of Pennsylvania, 209 South 33rd Street, Philadelphia, PA 19104}
\altaffiltext{15}{Department of Astronomy and Astrophysics, Pennsylvania State University, 525 Davey Laboratory, University Park, PA 16802, USA}
\altaffiltext{16}{The Oskar Klein Centre, Department of Astronomy, AlbaNova, Stockholm University, SE-106 91 Stockholm, Sweden}
\email{mathew.smith@uct.ac.za}

\begin{abstract}
Using data from the Sloan Digital Sky Supernova Survey-II (SDSS-II SN Survey), we measure the rate of Type Ia Supernovae (SNe Ia) as a function of galaxy properties at intermediate redshift. A sample of 342 SNe Ia with $0.05<z<0.25$ is constructed. 
Using broad-band photometry and redshifts we use the \PEGASE spectral energy distributions (SEDs) to estimate host galaxy stellar masses and recent star-formation rates. 
We find that the rate of SNe Ia per unit stellar mass is significantly higher (by a factor of $\sim 30$) in highly star-forming galaxies compared to passive galaxies. When parameterizing the SN Ia rate ($\snrIa$) based on host galaxy properties, we find that the rate of SNe Ia in passive galaxies is not linearly proportional to the stellar mass, instead a $\snrIa \propto {M}^{0.68}$ is favored. However, such a parameterization does not describe the observed SN Ia rate in star-forming galaxies. The SN Ia rate in star-forming galaxies is well fit by $\snrIa = 1.05 \pm 0.16 \times {10}^{-10} \mgal^{0.68 \pm 0.01} + 1.01 \pm 0.09 \times {10}^{-3} {\dot{M}^{1.00 \pm 0.05}}$ (statistical errors only), where $M$ is the host galaxy mass (in $\msun$) and $\dot{M}$ is the star-formation rate (in $\mperyr$). 
These results are insensitive to the selection criteria used, redshift limit considered and the inclusion of non-spectroscopically confirmed SNe Ia. 
We also show there is a dependence between the distribution of the MLCS light-curve decline rate parameter, $\Delta$, and host galaxy type. Passive galaxies host less luminous SNe Ia than seen in moderately and highly star-forming galaxies, although a population of luminous SNe is observed in passive galaxies, contradicting previous assertions that these SNe Ia are only observed in younger stellar systems. The MLCS extinction parameter, $A_V$, is similar in passive and moderately star-forming galaxies, but we find indications that it is smaller, on average, in highly star-forming galaxies. We confirm this result using the SALT2 light-curve fitter. 
\end{abstract}


\section{Introduction}
\label{sec:Introduction}

Type Ia supernovae (SNe Ia) have been extensively studied because they provide accurate relative distances on cosmological scales. Measurements of SNe Ia have indicated that the expansion of the universe is currently accelerating \citep{1998AJ....116.1009R, 2006A&A...447...31A, 2007ApJ...666..694W,2009ApJS..185...32K,2010MNRAS.401.2331L}, leading to the introduction of a ``Dark Energy" component in our model of the Universe. 

SNe Ia are thought to arise from carbon-oxygen white dwarfs that accrete mass from a companion star and approach the Chandrasekhar mass limit, resulting in a thermonuclear explosion \citep{1960ApJ...132..565H, 1995PASP..107.1019B, 1998ApJ...497..168Y}. However, there is still significant debate on the details; e.g. the explosion mechanism, the accretion process, and the progenitor companion star, which may be a giant star, a main sequence star, or a secondary white dwarf \citep{2003LNP...635..203H}. A measurement of the delay time (\textit{i.e.}, the time between the formation of the binary system and its thermonuclear explosion), constrains the possible progenitor systems \citep{2005A&A...441.1055G}.  The delay time distribution can be determined observationally by comparing the observed SNe Ia rates in galaxies with different star-formation histories \citep{2004MNRAS.347..942G}.

It has been observationally determined that SNe Ia are distinctly more common in galaxy hosts with recent star-formation activity \citep{1979AJ.....84..985O}.  Recent work has determined that the SNe Ia rate per unit stellar mass depends on host galaxy morphology and (B-K) color \citep{2005A&A...433..807M} and that the SN Ia rate in late-type galaxies is a factor $\sim20$ higher than in E/S0 galaxies. SNe Ia are seen locally to be rarer in galaxy bulges than spiral arms \citep{1997ApJ...483L..29W} and more common in blue galaxies than red \citep{2005A&A...433..807M}. The population associated with star-formation suggests that the SN Ia rate contains a population with a cosmologically short time delay, while the observation of SNe Ia in very old systems indicates the existence of a population with large time delay \citep{1999A&A...351..459C}.

\citet{2005ApJ...629L..85S} and later \citet{2006MNRAS.370..773M} and \citet{2006ApJ...648..868S} proposed a ``two-component" SN Ia rate, consisting of a prompt component, dependent on recent host galaxy star-formation, and a delayed component dependent on galaxy stellar mass. The overall SN Ia rate is thus the sum of these two components, and can be further generalised as a function of the galaxy star-formation rate and stellar mass. Observations strongly favor a two-component model over a single component model \citep{2006ApJ...648..868S} and since the cosmic star-formation rate increases with redshift, we expect that the prompt component will become a larger fraction of the SN Ia population with increasing redshift. \citet{2010ApJ...722.1879M}, using cluster rate measurements, suggest a universal delay time distribution, independent of environment and parameterized by $\snrIa \propto {t}^{-1.2 \pm 0.3}$.

Several attempts have been made to constrain the functional form of the SNe Ia rate. \citet{2006ApJ...648..868S}, using 124 SNe Ia from the SNLS survey, found that, for passive galaxies, the SNe Ia rate is consistent with a linear relationship with host galaxy stellar mass. Recently, \citet{2010arXiv1006.4613L}, used a sample of 274 SNe Ia from the LOSS survey, to consider how the size and morphology of the host galaxy affects the SNe Ia rate. They favor a power-law relationship between galaxy stellar mass and the SNe Ia rate (SNuM) with exponent approximately one half independent of both galaxy morphology and color. \citet{2010arXiv1002.3056M} find evidence for both a ``prompt" (age $< 420\, \textrm{Myr}$) and ``delayed'' component ranging between $2.4$ and $13\, \textrm{Gyr}$.\citet{2010arXiv1006.4613L} also show that SN Ia rate in young stellar populations may be strongly correlated with the rate of core-collapse SN.

A SN Ia rate composed of two components may have ramifications when SNe Ia are used to determine cosmological parameters. Since the two components are likely to have two different progenitor systems, the common assumption that all SNe Ia can be normalized in the same way is in question. It is likely that two distinct progenitor systems would contribute to the observed intrinsic scatter in the distances measured with SNe Ia. The relation between light-curve decline rate and peak luminosity for SNe Ia has been well tested \citep{1993ApJ...413L.105P, 2007ApJ...667L..37H}, but the physics behind this relation and the details of the explosion mechanism are only partially understood \citep{2007ApJ...656..661K}. Better understanding of the nature of SNe Ia explosions and progenitor systems will aid in improving the accuracy of SN Ia distance measurements \citep{2010arXiv1005.4687L,2010MNRAS.406..782S}. 

The total SN Ia rate has been well studied. \citet{2006AJ....132.1126N} and \citet{2011arXiv1102.0005G} have constrained on the SN Ia rate at high redshift, using the SNLS dataset to $z = 0.5$ and SNe Ia from the Subaru Deep Field (SDF) to $z < 2$, respectively.  \citet{2008ApJ...682..262D} determined the most accurate SNe Ia rate at intermediate redshift using the first year SDSS-II SN dataset to $z \le 0.12$ and extended the analysis to $z<0.3$ in \citet{2010ApJ...713.1026D}.

In this paper, we investigate the characteristics of the host galaxies of SNe Ia at intermediate redshift using the SDSS-II SN dataset following the methodology described in \citet{2006ApJ...648..868S}.  The SDSS-II SN survey is ideally suited to this task because it provides the largest, unbiased dataset currently available, with well understood efficiency corrections, in a redshift range fully in the Hubble flow, but with high signal-to-noise observations. Our host galaxies are well measured in several filter bands, allowing us to accurately estimate the galaxy's properties. The aim of this work is to measure the rate of SNe Ia explosions as a function of the host galaxy stellar mass and star-formation rate. We parameterize the relationship for our intermediate redshift data, and interpret the results in the context of the two-component model.

This paper is organized as follows. In \S\ref{sec:SDSS} we describe the SDSS-II SN Survey, the observing strategy and give a brief account of the results of this survey. In \S\ref{sec:incompleteness} we show how the SDSS-II SN dataset is incomplete and introduce a method to produce an unbiased sample with a well understood efficiency correction. In \S\ref{sec:galaxydata} we outline how the host galaxy of each SNe Ia is identified and the method that we use to determine the derived properties of the host galaxy. We also describe the sample of field galaxies used for comparison, and how it is corrected for incompleteness. In \S\ref{sec:hostrate} we investigate how the SN Ia rate is dependent on the properties of the host galaxy, studying how it is dependent on the stellar mass of the galaxy (\S\ref{sec:massrate}), the star-formation rate of the host galaxy (\S\ref{sec:sfrrate}), and the specific star-formation rate of the host (\S\ref{sec:ssfr}). Finally, \S\ref{sec:sneprop} discusses how the light-curve shape (\S\ref{sec:snprop_delta}) and extinction (\S\ref{sec:snprop_av}) of the SN Ia events are related to the galaxy host properties. Our conclusions are given in \S\ref{sec:conclusions}.

%%%%%%%%%%%%%%%%%%%%%%%%%%%%%%%%%%%%%%%%%%%%%%%%%%%%%%%%%%%%%%%%%%%%%%%%%%%%%%%%%%

\section{The SDSS-II SN Survey}
\label{sec:SDSS}

In this work, we use the full sample from the SDSS-II SN Survey \citep{2008AJ....135..338F}. This provides one of the largest samples of SNe Ia currently available. 

The SDSS-II SN Survey was a three year rolling search that produced a sample of spectroscopically confirmed SNe Ia with well-measured multi-color light-curves at intermediate redshift ($z<0.4$) using the SDSS 2.5m telescope \citep{2000AJ....120.1579Y,2002AJ....124.1810S,2006AJ....131.2332G} at Apache Point Observatory with a wide field CCD camera \citep{1998AJ....116.3040G}. Observations were made in the SDSS \textit{ugriz} filters \citep{1996AJ....111.1748F}, alternating between the northern and southern ``strips" of the  field designated as ``Stripe 82" \citep{2002AJ....123..485S}, bounded by $-60 \degree < \alpha (\textrm{J} 2000) < 60 \degree$, and $-1.258 \degree < \delta (\textrm{J} 2000) < 1.258 \degree$.  Adverse weather and bright moonlight resulted in an average observation of each strip once every four nights with typical limiting magnitudes of  $\textit{g} \sim 21.8$, $\textit{r} \sim 21.5$, $\textit{i} \sim 21.2$ per observation. The scene modelling photometry (SMP) technique of \citet{2008AJ....136.2306H} was used to produce accurate photometric data for each SN event. 

The SDSS-II SN Survey identified many thousands of transient events, of which $513$ were spectroscopically confirmed as SNe Ia  and $85$ were other SN types \citep{2008AJ....135..348S,2008AJ....135.1766Z}.  Spectroscopic redshifts for the host galaxies of $339$ probable SNe Ia, based on their light-curves, were also obtained, and are discussed further in \S\ref{sec:incompleteness}. 

The first year SDSS-II SN sample was used for a cosmological analyses \citep{2009ApJS..185...32K,2009ApJ...703.1374S,2010MNRAS.401.2331L}. \citet{2008ApJ...682..262D,2010ApJ...713.1026D} measured the SNe Ia volumetric rate, \citet{2010arXiv1005.4687L,Gupta2011,Konishi2011} and \citet{DAndrea2011} analyzed the effect of host galaxies on light-curve parameters, both from the photometric properties of the host galaxies and studying their spectral features.

The SDSS-II SN Survey is approximately magnitude limited, producing an otherwise unbiased sample. This analysis uses a sample of 342 SNe Ia in the redshift range $0.05<z<0.25$, where the efficiency of the survey is high \citep{2010ApJ...713.1026D}. This homogenous sample is comprised of 197 spectroscopically confirmed SNe Ia, with a further 87 having a host galaxy spectroscopic redshift. All objects are selected using a well defined selection criteria, and have well-measured light-curves that are consistent with a SNe Ia template, based on the Bayesian light-curve fitting method of \citet{2008AJ....135..348S}. The selection criteria used to create this sample is discussed in \S\ref{sec:incompleteness}. 

%%%%%%%%%%%%%%%%%%%%%%%%%%%%%%%%%%%%%%%%%%%%%%%%%%%%%%%%%%%%%%%%%%%%%%%%%%%%%%%%%%

\section{Incompleteness Corrections}
\label{sec:incompleteness}

There are two major sources of inefficiency in the SDSS-II SN pipeline that lead to potential biases in the spectroscopically confirmed SN sample: detection efficiency and spectroscopic incompleteness. The detection efficiency was primarily magnitude limited and is amenable to calculation by simulation.  The spectroscopic selection and analysis depends on many factors that are difficult to quantify. We adopt a strategy of augmenting the sample of spectroscopically confirmed SNe Ia with a sample of photometrically classified SNe Ia, identified by their light-curve shape and color and correcting for detection efficiency.

\subsection{Correcting for Spectroscopic Incompleteness}
\label{sec:spectroissues}

The SDSS-II SN Survey prioritized spectroscopic follow-up observations of SN candidates using a Bayesian classification method \citep{2008AJ....135..348S}\footnotemark. However, the final ranking and decisions on spectroscopic follow-up priorities were based on the telescope's capabilities, local weather conditions and the SN position on the sky, thus leading to a spectroscopic sample whose selection criteria are difficult to describe quantitatively. To produce a homogeneous sample of SN Ia candidates, we therefore seek a sample selection that avoids the uncertain and time-varying spectroscopic target selection process.  We also seek a sample with high quality light-curves and low levels of contamination. However, we must also ensure that the majority of detected SNe Ia pass this criteria, so that our results are not dominated by the efficiency corrections. 
\footnotetext{An updated version of this method is given in \citet{Sako2011}. However, as our goal is to replicate the follow-up strategy of the SDSS-II SN Survey, it is not used in this work.}

We adopt a two-stage process. In the first stage, we use photometry obtained during the SN Ia search and the Bayesian classification method, to apply very loose cuts that are intended to reduce the large number of non-SN Ia transient objects that are classified as candidates by the SDSS-II search pipeline, while retaining any SN Ia that could possibly survive our subsequent quality cuts. This sample is then analyzed by the more accurate SMP photometry and fit by the MLCS2k2 light-curve fitter to obtain a sample of probable SN Ia. The criteria used for our two-stage process is described as follows.

Firstly, as part of the SDSS-II SN operations, every transient object with more than two epochs was selected to be a candidate, after known AGNs, variable stars and pipeline artifacts were removed. There are $\sim20,000$ such candidates. The Bayesian classification technique, used in the SDSS-II SN search operations, fits SNe Ia, Ib/c and II template light-curves to each candidate, producing a probability, $p_T$, for a candidate to belong to each class ($T$) of SNe. This method assumes that each candidate is a SN of some particular type, but has been shown, nevertheless, to be accurate in differentiating between different SN types \citep{2008AJ....135..348S,2010PASP..122.1415K}. This Bayesian classification technique was applied to each candidate, and the following criteria was used to select viable SN Ia candidates:
\begin{itemize}
\item At least 3 search discovery epochs,
\item $p_{Ia} > 0.45$,
\item If the candidate has more than 5 search photometry epochs, the best-fit Ia model is not SN 2005gj\footnotemark.
\end{itemize}
\footnotetext{SN 2005gj \citep{2006ApJ...650..510A,2007arXiv0706.4088P} is a peculiar SN, with a flat light-curve after maximum.  In addition to removing SN 2005gj-like SNe, this criterion also removes AGN and other non-transient events from our sample. }

Additional cuts were considered, including using the photometric redshift from the nearest host galaxy to constrain the light-curve, but were rejected, as it is significantly harder to model the SDSS-II SN survey selection function with those cuts. Our criteria select $1762$ candidates, including $88\%$ of the spectroscopically confirmed SNe Ia. Of the $12\%$ of confirmed SNe Ia that fail this selection criteria, $27\%$ (17) were only observed on one or two occasions, $70\%$ (45) do not satisfy the $p_{Ia}$ criteria, and $3\%$ (2) are best-fit by a 2005gj-like template (including SN 2005gj itself).

The selection criteria described above, uses photometry obtained during the SN Ia search to produce a sample of candidates containing the vast majority of the spectroscopically confirmed SNe Ia, whilst removing the vast majority of non-SN Ia transient objects. In the second stage of our selection criteria, this sample was then analyzed using the more complete and more accurate SMP photometry and fit using the MLCS2k2 light-curve fitter \citep{2007ApJ...659..122J,2009PASP..121.1028K} to ensure each candidate has a well covered light-curve and is well fit by an SN Ia event. The selection criteria are the same as were used by \citet{2009ApJS..185...32K} and \citet{2010ApJ...713.1026D}, namely,
\begin{enumerate}
\item At least 5 photometric observations (all at different epochs) between $-20$ and $+60$ days relative to peak light in the rest-frame of the SN,
\item At least one epoch with signal-to-noise ratio $> 5$ in each of \textit{g}, \textit{r}, and \textit{i} (not necessarily the same epoch in each passband),
\item At least one photometric observation at least 2 days prior to maximum brightness in the SN rest frame,
\item At least one photometric observation at least 10 days past maximum brightness in the SN rest frame,
\item MLCS2k2 light-curve fit probability $> 0.001$\footnotemark, 
\footnotetext{ 6 of the spectroscopically confirmed SNe Ia fail this criteria, including 4 peculiar SN Ia}
\item MLCS2k2 light-curve decline rate parameter of $\Delta > -0.4$ \footnotemark,
\item $-51 \degree < \alpha (\textrm{J} 2000) < 57 \degree$.
\end{enumerate}
\footnotetext{The cuts on MLCS2k2 light-curve fit probability and $\Delta$ (5,6) have a negligible effect on the size of our sample compared to the sampling cuts (1,2,3,4)}

Excess color in SNe Ia is interpreted by MLCS2k2 as extinction by dust in the host galaxy, parameterized using \citet{1989ApJ...345..245C}, where  $E(B-V) = A_V / R_V$.  For this analysis, we adopt a value of $R_V=2.3$ and assume an $A_V$ prior in the fitting process of $P\left(A_V \right) = e^{-A_V / \tau}$, with $\tau=0.33$, as described in \citet{2009ApJS..185...32K}. For comparison, $R_V=3.1$ on average for our galaxy, but previous SN Ia studies have favored values of $R_V \sim 2.0$ \citep{2008A&A...487...19N,2010arXiv1005.4687L}. 

Of the 1762 candidates that satisfy the Bayesian light-curve fitter criteria, 843 satisfy the sample selection. Of these 843, 319 are spectroscopically confirmed as SNe Ia and 180 are unconfirmed but have a host galaxy spectroscopic redshift.  

The SNANA version \citep{2009PASP..121.1028K} of MLCS is able to estimate a photometric redshift for SN candidates in addition to determining a distance modulus. Most of our 843 candidates lack spectroscopic redshift measurements, so we adopt a cosmological model of $\omatter = 0.3$ and $\olambda = 0.7$, to reduce the number of fit parameters, and to determine a photometric redshift for each candidate. To construct a sample that is unbiased with respect to spectroscopic follow-up, and has a well determined selection function, we fit for a photometric redshift for all candidates, regardless of whether a spectroscopic redshift is known. An analysis of the accuracy of these photometric redshift estimates is given in \citet{2010ApJ...713.1026D}, who find that the photometric redshifts are negligibly biased and are accurate to $\sim 0.01$ at low redshift $(0 < z < 0.25)$

These selection criteria ensure that each candidate has a well covered light-curve that is well fit by a normal SN Ia event (peculiar SNe Ia will generally not pass the selection criteria). 

While we have relied on photometric redshifts for the initial sample selection, we use spectroscopic redshift information, when available, for redshift selection and all subsequent analysis.  To construct a sample that is primarily comprised of spectroscopically confirmed SNe Ia and not dominated by photometrically classified SNe Ia, and to avoid low detection efficiency (see \S\ref{sec:efficiency}), we restrict our SN sample to the redshift range $0.05 < z < 0.25$. This leaves 379 SNe Ia. We find 217 ($57\%$) are spectroscopically confirmed SNe Ia, while 94 ($25\%$) are unconfirmed but have host galaxy spectroscopic redshifts and 68 ($18\%$) have no spectroscopic redshift information.  The number of candidates that satisfy each stage of our selection criteria is shown in Table~\ref{tab:selection}. While the majority of our sample has been spectroscopically confirmed, a significant fraction of candidates are only photometrically classified. However, \citet{2010ApJ...713.1026D} conservatively estimated that there is a $3\%$ probability for non-SNe Ia to satisfy our selection criteria, and the total estimated contamination by non-SNe Ia's is $2\%$.

\begin{deluxetable}{ccc}
\tabletypesize{\scriptsize}
\tablecaption{Number of candidates passing each stage of \S\ref{sec:spectroissues} \label{tab:selection}}
\tablewidth{0pt}
\tablehead{ \colhead{} & \colhead{No. of Candidates} & \colhead{Spectroscopically confirmed} }
\startdata
All SDSS-II SN candidates & 19046 & 513 \\
After Bayesian LC fit & 1762 & 449 \\
Passing Sample Selection & 843 & 319 \\
$0.05 < z < 0.25$ & 379 & 217 \\  
\enddata
\end{deluxetable}

Table~\ref{tab:noofsne} lists the number of SNe Ia that pass our selection criteria for several redshift ranges, including the proportion of each sample that is spectroscopically confirmed as SNe Ia. As expected, the proportion of spectroscopically confirmed SNe decrease with increasing redshift, but it remains above $50\%$ out to $z=0.25$. 
\begin{deluxetable}{ccccc}
\tabletypesize{\scriptsize}
\tablecaption{Number of candidate SNe Ia as a function of redshift \label{tab:noofsne}} 
\tablewidth{0pt}
\tablehead{ \colhead{Redshift Limit} & \colhead{Total} & \colhead{Spectroscopically confirmed} & \colhead{Host Redshift} & \colhead{Photo-z only} }
\startdata
$0.05 < z < 0.10$ & 21& 19 ($90.5\%$) & 1 ($4.8\%$) & 1($4.8\%$) \\
$0.05 < z < 0.15$ & 88 & 73 ($83.0\%$) & 11 ($12.5\%$) & 4 ($4.5\%$) \\
$0.05 < z < 0.20$ & 214 & 144 ($67.3\%$) & 47 ($22.0\%$) & 23 ($10.7\%$) \\
$0.05 < z < 0.25$ & 379 & 217 ($57.3\%$) & 94 ($24.8\%$) & 68 ($17.9\%$) \\
$0.05 < z < 0.30$ & 559 & 272 ($48.7\%$) & 141 ($25.2\%$) & 146 ($26.1\%$) \\
$0.05 < z < 0.40$ & 800 & 312 ($39.0\%$) & 176 ($22.0\%$) & 312 ($39.0\%$) \\
\enddata
\end{deluxetable}

\subsection{Determining the Survey Efficiency}
\label{sec:efficiency}

\begin{figure}[ht]
\epsscale{1.0}
\plotone{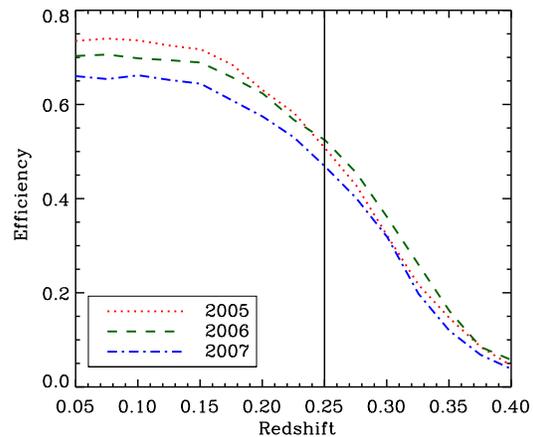}
\caption{The efficiency of SN detection as a function of redshift for each observing season. The high redshift limit used in this analysis is also shown. \label{fig:efficiency}}
\end{figure}

Having defined a homogeneous sample of SNe Ia candidates with $0.05 < z < 0.25$, we need to know the SDSS-II SN detection efficiency, $\epsilon(z)$. A detailed analysis of the efficiency was given in \citet{2008ApJ...682..262D,2010ApJ...713.1026D}, differing here only in the use of MLCS fitted photometric redshifts to select the fake SNe that pass our redshift cut. Simulated SNe Ia, with a range of sky positions, time of peak brightness, redshifts, decline-rate parameters, extinction and host galaxy position, and realistic errors were added directly to the image data and were processed by the SDSS-II SN pipeline \citep{2008AJ....135..348S}. The proportion of SNe Ia that satisfy the criteria defined in \S\ref{sec:spectroissues} is shown as a function of redshift in Figure~\ref{fig:efficiency} for each of the three years of the SDSS-II SN Survey. We also highlight the high redshift limit used in this analysis. Over $50\%$ of SNe Ia are detected in our redshift range.  The fact that the efficiency is less than $100\%$ at low redshift is caused by SN explosions that occur late or early in the observing season and fail to allow the required number of observations.  This inefficiency is a major effect at all redshifts but is accurately modelled in our simulation. The uncertainty on the survey efficiency is discussed in detail in \citet{2010ApJ...713.1026D}. 

%%%%%%%%%%%%%%%%%%%%%%%%%%  Figure 1  %%%%%%%%%%%%%%%%%%%%%%%%%%%%%%
%%%%%%%%%%%%%%%%%%%%%%%%%%%%%%%%%%%%%%%%%%%%%%%%%%%%%%%%%%%%%

The survey efficiency considered for this analysis is a function of redshift. However, this functional form maybe too simplistic, as it assumes no variation in the intrinsic brightness of SNe Ia as a function of host galaxy type. To determine if our results are dependent on this assumption, we study how our conclusions are affected if parameterize the survey efficiency as a function of redshift, $A_V$ and $\Delta$. This additional correction produces results that are consistent with our nominal result, with differences of much less than $1\sigma$. We thus consider the survey efficiency to be a function solely of redshift, but note that other analyses have not considered the effect of this assumption on their results.

We have now defined a uniformly selected sample of 379 SNe Ia candidates with $0.05 < z < 0.25$ from the three years of the SDSS-II SN Survey, and calculated the efficiency of the survey in this redshift range, which we shall invert to weight the galaxies in our sample. The uncertainty on the survey's efficiency is small \citep{2010ApJ...713.1026D} compared to the statistical precision of our data and it is not necessary for us to include the uncertainty in our analysis. We now turn to consider the host galaxies of these SN events.  

%%%%%%%%%%%%%%%%%%%%%%%%%%%%%%%%%%%%%%%%%%%%%%%%%%%%%%%%%%%%%%%%%%%%%%%%%%%%%%%%%%

\section{Host Galaxy Determination and Derived Quantities}
\label{sec:galaxydata}

Here we describe the method used to identify the host galaxies and determine their characteristics, such as stellar mass and recent star-formation rate, for the 379 SNe Ia identified in \S\ref{sec:incompleteness}. We also outline the comparison field sample used to describe the underlying galaxy population in our redshift range. 

\subsection{Host Galaxy Determination}
\label{subsec:determiningthehost}

Repeat imaging of SDSS Stripe 82 has enabled the coaddition of images into a deep stacked image \citep{2009ApJS..182..543A}. The stack ranges from 20 to 40 individual images (depending on sky position) in all 5 SDSS filters (\textit{ugriz}) and is roughly 2 magnitudes deeper than a single epoch SDSS image. To determine the host galaxy for each SN in our sample, we match the SN positions with SDSS galaxies detected in this deep stacked image within a $0.25$ arcminute radius. We require that the host galaxy has an SDSS model magnitude \citep{2002AJ....123..485S} in the range $15.5 < \textit{r} < 23.0$ to ensure robust photometry. This magnitude cut is conservative, but applied to ensure that the SDSS-II SN pipeline is able to accurately distinguish between stars and galaxies in the deep stacks. The magnitude limits remove $10\%$ of our SNe with either unobserved or too faint hosts. We then visually scan each host galaxy, using images with and without the supernova present to ensure that our host galaxy association is accurate. In six cases, at low redshift, where the host is extended or resolved into multiple objects, we select by hand a more likely object as the host galaxy.  Of the 379 SNe Ia candidates identified in \S\ref{sec:incompleteness}, 342 have a valid host galaxy identification, of which 197 ($58\%$) are spectroscopically confirmed to be SNe Ia, and 87 ($25\%$) are spectroscopically unconfirmed but have a host galaxy spectroscopic redshift. The remaining 58 objects are classified to be SNe Ia through their photometry alone. Of the 37 candidates that lack a valid host galaxy, 29 ($78\%$) have a host galaxy candidate with $\textit{r} > 23.0$ and 8 ($22\%$) have no host candidate within a $0.25$ arcminute radius.

\subsection{Derived Host Galaxy Properties}
\label{subsec:galprop}

Having identified the host galaxy position and magnitudes for 342 SNe Ia candidates, we now determine their stellar mass and recent star-formation rate. 

There are several methods to infer galaxy properties from broad-band photometry. A simple cut on the color of the galaxy can be used to infer its spectral type \citep{2001AJ....122.1861S} and the UV flux can provide an estimate of the recent star-formation rate \citep{1987A&A...180...12D}. These simple methods are able to differentiate between galaxies with markedly different levels of star-formation activity, but struggle with galaxies with similar colors because multi-band photometry is not used \citep{2006MNRAS.373..469B}. Therefore, we fit our multi-band photometry to a set of Spectral Energy Distributions (SEDs) and use the best-fit template to determine the galaxy parameters. This technique is widely used for photometric redshift estimates \citep{2000A&A...363..476B,2002A&A...386..446L,2008ApJ...674..768O}.

\subsubsection{SED Fitting}
\label{subsubsec:galfitting}

The method used here is consistent with that of \citet{2006ApJ...648..868S}, who studied the SN Ia rate as a function of host galaxy properties at high redshift, allowing our results to be compared within the same framework. A discussion on how the different redshift ranges covered by this analysis and that of \citet{2006ApJ...648..868S} may affect our host galaxy derived properties is given in Appendix~\ref{app:restwave}. 

We use the SEDs produced by the \PEGASE galaxy spectral evolution code \citep{1997A&A...326..950F,2004A&A...425..881L}. These templates have been used extensively in the literature to constrain the evolution of galaxies, particularly at high redshift \citep{2004Natur.430..181G,2006A&A...449..951G}. We use the set of 8 evolutionary tracks listed in Table 1 of \citet{2002A&A...386..446L} (excluding the starburst template), and assume a \citet{2001MNRAS.322..231K} Initial Mass Function (IMF). In these scenarios, star-formation rate is determined using the relationship $\textrm{SFR} = \nu \times M_{\textrm{gas}}$, where $\nu$ ranges from 0.07 to $3.33\, \textrm{Gyr}^{-1}$, and $M_{\textrm{gas}}$ is the density of gas in solar masses. Extinction due to dust is modelled internally, with a \citet{1980ApJ...241..474K} profile used for the Elliptical template, and a plane-parallel slab geometry is used for the spiral and irregular templates. Each of the 8 evolutionary scenarios is evolved over 69 time steps, each one corresponding to a different galaxy age, making a total of 552 template SEDs. 

These SEDs are convolved with the SDSS filter responses \citep{1996AJ....111.1748F} and fitted to the galaxy fluxes (calculated from model magnitudes after correcting for Galactic dust absorption from \citet{1998ApJ...500..525S} and AB-system offsets) using the Z-PEG photometric redshift code \citep{2002A&A...386..446L}. We keep the redshift of the SN host galaxies fixed to the spectroscopic redshift (from either the SN or host galaxy) or the photometric redshift determined by MLCS2k2.  Applying a redshift constraint eliminates the color uncertainty due to the cosmological redshift.  As dust is included internally in the SEDs no dust correction is applied in the fitting process. We assume a default $\Lambda$CDM cosmology ($\omatter = 0.3, \olambda = 0.7$) and consider only templates that are younger than the age of the Universe at the fitted redshift. 

The best-fit template is determined through a  $\chi^2$ minimization using all 5 SDSS filters. The total stellar mass of each galaxy is determined by integrating the star-formation history of the best-fitting SED and subtracting the mass of stars that have died. We characterize recent star-formation with a mean star-formation rate, since the instantaneous star-formation rate is difficult to estimate without high-resolution spectroscopic data. We use the result of \citet{2006ApJ...648..868S}, who found that averaging the star-formation rate over a period of $0.5\,\textrm{Gyr}$ can be accurately recovered by the \PEGASE SEDs, without introducing significant systematic uncertainties, especially for galaxies where the redshift is unknown. 

Uncertainties in the galaxy properties are determined from the range spanned by the SEDs satisfying $\chi^2 \leq {\chi}^2_{\textrm{min}} + 1$. We consider errors on the galaxy fluxes from the coadded image, with a minimum error as given in \citet{2007AJ....133..734B}. The stellar mass and recent star-formation rate for the 342 host galaxies used in this analysis is given in Table~\ref{tab:listofobjects}. 

\subsection{Comparison Field Sample}
\label{subsec:fieldsample}

To determine how our SN sample relates to the underlying galaxy population in our redshift range, we require a sample of galaxies that is representative of the general galaxy population. For this sample, we use galaxies detected in the deep stacks described in \S\ref{subsec:determiningthehost}. We consider galaxies identified in the SDSS-II SN Survey region with $15.5 < r < 23.0$. Thus cut also removes the possibility of variable limiting magnitudes across the image. 

We determine the stellar masses and recent star-formation rates for each galaxy in this sample using the same method as for the host galaxy sample except that the redshift is a free parameter to be determined by the Z-PEG fit. We require that the fitted redshift must lie in the redshift range $0 < z < 2$. The additional freedom allowed in determining the redshift for the field galaxies can result in large error bars on the derived photometric redshift, stellar mass and star-formation rate estimates.  In extreme cases, there can be two or more distinct best-fit template solutions, resulting in more than one photometric redshift estimate and spectral type. In these cases, the galaxy is excluded from our analysis because the spectral classification and derived galaxy properties are ambiguous. To match the host galaxy population, we consider the $\sim750,000$ galaxies with $0.05 < z < 0.25$. 

\subsection{Correcting For Incompleteness in the Field Sample}
\label{subsec:fieldincompleteness}

The comparison field sample is magnitude limited, and thus becomes increasingly incomplete at higher redshifts, with only the brightest galaxies observed at higher redshifts. Galaxies with a given absolute magnitude (and spectral type) will pass the apparent magnitude selection criteria ($15.5 < r < 23.0$) at different redshifts, which may be less than the full survey range ($0.05 < z < 0.25$). To correct for this effect, we use the $V_\textrm{max}$ method \citep{1968ApJ...151..393S,1976ApJ...207..700F}. Using the best-fitting SED for each field galaxy, we calculate its absolute magnitude and k-correction, and determine the redshift limits at which it would satisfy $15.5 < r < 23.0$.  Whenever the redshift range is less than the total survey range ($0.05 < z < 0.25$), we weight the galaxy by ${V}_{\textrm{survey}} / {V}_{\textrm{max}}$, where ${V}_{\textrm{max}}$ is the co-moving volume for which each galaxy will remain within our survey's magnitude limits, and ${V}_{\textrm{survey}}$  is the co-moving volume of the SDSS-II SN survey, \textit{i.e.} for a redshift range, $0.05 < z < 0.25$ and constant for each galaxy in our sample. $83\%$ of the field galaxies in our sample have redshift limits larger than that of the SDSS-II SN survey, and are not affected by this correction. The remaining $17\%$ of field galaxies are on average weighted by a value of $4.98$. Since this form of incompleteness will affect both the comparison field sample and host galaxy sample, this incompleteness correction is applied to both, although only 3 of the 342 host galaxies in our sample are affected by this correction. 

\subsection{Systematic Uncertainties}
\label{subsubsec:systematics}

Systematic uncertainties in our derived galaxy properties can arise from many sources including the wavelength coverage of the SDSS filters, our decision to use the \PEGASE SEDs, our choice of IMF, the accuracy of the photometric redshifts for the comparison field sample, the accuracy of the \PEGASE stellar mass estimates, and the ability of \PEGASE to accurately recover the stellar masses and star-formation rates for a sample of simulated galaxies. All these systematic errors are discussed further in Appendix~\ref{app:color-mag},~\ref{app:pegase_photoz},~\ref{app:effect},~\ref{app:brinc} and~\ref{app:restwave}.

In Appendix~\ref{app:color-mag}, we show that the \PEGASE SEDs primarily use the color of a galaxy as a proxy to infer its spectral type. The reddest galaxies are classified as passive galaxies, with the bluest galaxies considered highly star-forming. Moderately star-forming galaxies are distributed between passive and highly star-forming galaxies, spanning a large range of color. 

In Appendix~\ref{app:pegase_photoz}, we investigate the accuracy of the \PEGASE photometric redshift estimates for our field sample. We find a mean offset in redshift of $\Delta z = 0.03$, with the photometric redshift estimate being smaller than the known spectroscopic redshift.  This redshift error results in an error in stellar mass of $\Delta \log \mgal = 0.22\, \msun$. In Appendix~\ref{app:effect} we show the effect that applying this offset to our data would have on the results presented in \S\ref{sec:hostrate} and show that they are consistent. This offset provides us with an estimate of our systematic uncertainty, but due to a lack of understanding of the cause of this offset, it is not applied to our nominal analysis. 

Appendix~\ref{app:brinc} studies how the stellar mass and star-formation rate estimates from \PEGASE for our host galaxy sample compare to those determined using the spectral features of galaxies. We consider a sample of SDSS galaxies that have spectroscopically measured stellar masses and star-formation rates \citep{2004MNRAS.351.1151B,2003MNRAS.341...33K} and compare these to estimates determined in our analysis. We find that the stellar masses are recovered, with no significant offset, but there is a mean offset in the star-formation rate of $\Delta \log \textrm{SFR} = 0.12\,\mperyr$.  However, \citet{2004MNRAS.351.1151B} measure the ``instantaneous" (present day) star-formation rate instead of our ``recent" star-formation rate, which is averaged over the last 0.5 Gyr, so the two quantities are not directly comparable. 

In Appendix~\ref{app:restwave} we consider how the rest-wavelength coverage of the SDSS filter set affects our stellar mass estimates. With increasing redshift, the SDSS filters will sample a different rest wavelength ranges. This can be particularly important for systems with a variety of stellar populations, such as merging galaxies. To examine the sensitivity to our wavelength coverage, we repeat the determination of stellar masses and star-formation rates using only three or four of the five SDSS filters. We find an increased scatter in the results, but no overall bias in the stellar mass or star-formation rate estimates. This is particularly encouraging, because it suggests that the comparison of our galaxy properties with those of \citet{2006ApJ...648..868S} will not be affected by the different cosmological redshifts of the two surveys.

%%%%%%%%%%%%%%%%%%%%%%%%%%%%%%%%%%%%%%%%%%%%%%%%%%%%%%%%%%%%%%%%%%%%%%%%%%%%%%%%%%

\section{Host Galaxy Properties}
\label{sec:hostgalprop}

In \S\ref{sec:incompleteness} we defined a sample of homogeneously selected SNe Ia, and, in \S\ref{sec:galaxydata}, determined a host galaxy for each object. Having estimated their stellar mass and recent star-formation rate, we now analyze these derived properties, and how they relate to the supernova rate. 

\begin{figure*}[t]
\epsscale{1.5}
\hspace{-1.0cm}\plotone{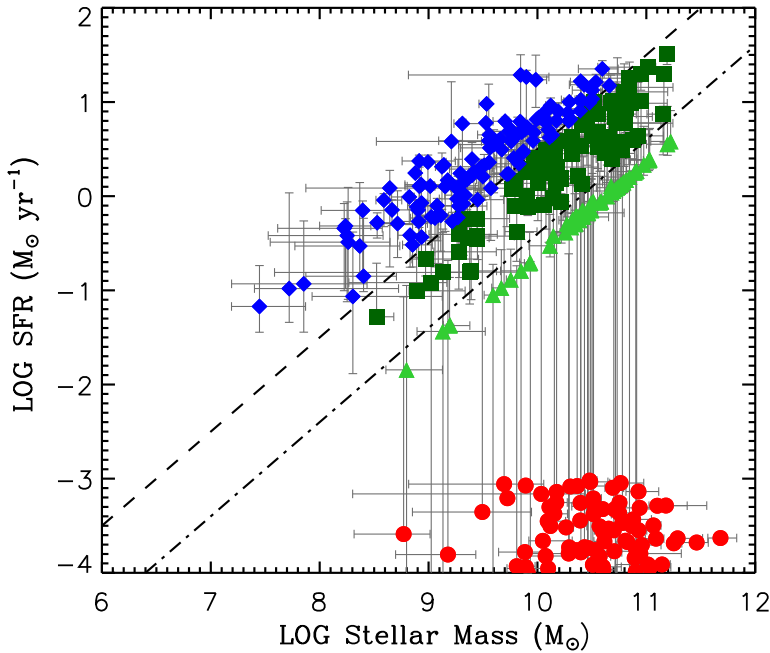}
\caption{The distribution of stellar mass \& $\mathrm{SFR}$ for the 342 SN host galaxies. Highly star-forming galaxies are shown as blue diamonds and passive galaxies as red circles. Moderately star-forming galaxies are plotted in green, with light green triangles indicating the ``ridge line" of galaxies discussed in the text and Appendix~\ref{app:color-mag}, and the remaining population plotted as dark green squares. The dashed-dotted line highlights this split. Passive galaxies have $\mathrm{SFR}=0$  but are shown here as randomly distributed in the range  $-4 < \log \mathrm{SFR} < -3$. The dashed line indicates the split used to distinguish highly star-forming galaxies from those with moderate levels of star-formation activity. .\label{fig:massvsfr}}
\end{figure*}

Figure~\ref{fig:massvsfr} shows the distribution of our host galaxy sample in stellar mass and star-formation rate (SFR).  Galaxies are shown in three categories, highly star-forming (blue), moderately star-forming (green), and passive (red). The highly and moderately star-forming galaxies are separated by their specific star-formation rate (sSFR): the star-formation rate per unit stellar mass \citep{1997ApJ...489..559G,2000ApJ...536L..77B,2004MNRAS.351.1151B,2006ApJ...648..868S}.  We followed  \citet{2006ApJ...648..868S} in choosing $\log \mathrm{sSFR} = -9.5$ as the arbitrary division between highly and moderately star-forming as indicated by the dashed line on Figure~\ref{fig:massvsfr}. Highly star-forming galaxies are using a significant proportion of their stellar mass to form new stars and their stellar populations are expected to be dominated by young, massive stars. Galaxies classified as moderately star-forming are likely to be dominated by an older, more evolved population of stars. Passive galaxies have a nominal $\mathrm{SFR}=0$, but for display purposes, are randomly distributed in red on Figure~\ref{fig:massvsfr} around $\log \mathrm{SFR}=-3.5$.  The average stellar mass of a passive galaxy is $\log M = 10.52\, \msun$, considerably more massive than star-forming galaxies, which average $\log M = 9.91\, \msun$, consistent with other observations of the local universe \citep{2009ApJ...694.1171T}. 

%%%%%%%%%%%%%%%%%%%%%%%%%%  Figure 2  %%%%%%%%%%%%%%%%%%%%%%%%%%%%%%
%%%%%%%%%%%%%%%%%%%%%%%%%%%%%%%%%%%%%%%%%%%%%%%%%%%%%%%%%%%%%%

Of the 342 galaxies in our sample, 80 ($23\%$) are classified as passive galaxies, 139 ($41\%$) have moderate levels of star-formation activity and 123 ($36\%$) are highly star-forming. 

In Figure~\ref{fig:massvsfr}, we note a ``ridge line" of galaxies, which are classified as moderately star-forming, but have the lowest possible values of sSFR allowed. A dashed-dotted line is shown on Figure~\ref{fig:massvsfr} to highlight this population of galaxies. $78\%$ of these galaxies are best-fit by the lenticular S0 (scenario), with the remaining $22\%$ being best-described by the elliptical template. In comparison $52\%$ of the remaining moderately star-forming galaxies are best-fit by the S0 scenario. In Appendix \ref{app:color-mag} we show the color-magnitude diagram, and conclude that these galaxies lie at the edge of the distribution of the moderately star-forming galaxies but appear to be distinct from the passive galaxies. Thus, we do not remove these galaxies from our analysis.  We will show later through a Monte-Carlo approach that removing these galaxies from our sample do not affect our major conclusions. 

%%%%%%%%%%%%%%%%%%%%%%%%%%%%%%%%%%%%%%%%%%%%%%%%%%%%%%%%%%%%%%%%%%%%%%%%%%%%%%%%%%

\section{SN Ia Rate}
\label{sec:hostrate}

We now turn to looking at how the supernova rate depends on the galaxy properties of total stellar mass and recent star-formation for passive and star-forming galaxies. 

\subsection{SN Ia Rate as a Function of Host Galaxy Stellar Mass}
\label{sec:massrate}

According to the standard model of galaxy formation, passive galaxies are primarily comprised of old, low mass stellar systems that evolve without forming new stars. It is reasonable to suppose that the SN Ia population in passive galaxies could only occur as a result of a process with a delay time that is long compared to the age of the galaxy.  If that is the case, then the number of SN Ia's occurring in these environments could be expected to be proportional to the host galaxy stellar mass. On the other hand, if the delay time is only comparable to the age of the galaxy, there could be a more complicated dependence based on the details of the star-formation history. 

\begin{figure*}[ht]
\epsscale{1.5}
\plotone{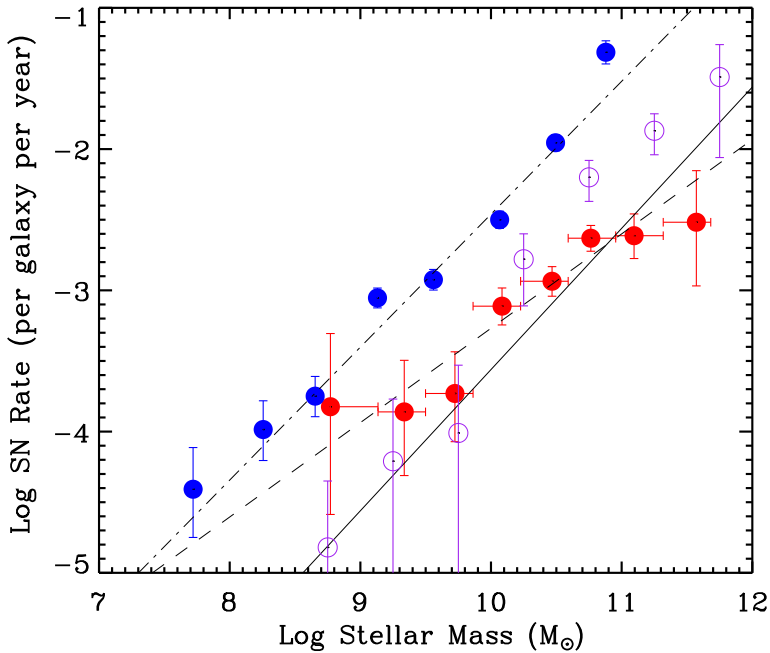}
\caption{SN Ia rate as a function of host galaxy stellar mass. The values for star-forming (passive) galaxies are shown in blue (red). The data points, for passive galaxies, from \citet{2006ApJ...648..868S} are shown as open circles. The best-fitting lines for passive and star-forming galaxies are shown as dashed and dotted-dashed lines, respectively. Also shown is a fit (solid line) to the passive galaxies where the line slope is assumed to be one. \label{fig:massrate}}
\end{figure*}

To measure the stellar mass dependence with the SDSS data, we split both our host galaxy sample and comparison field sample into passive and star-forming galaxies. The samples are binned by their stellar mass, with both the host galaxy and field sample weighted for incompleteness using the $1/V_\textrm{max}$ correction, and the efficiency correction applied to the host galaxy sample. Each host galaxy is weighted by $1/\epsilon$, the survey efficiency at the redshift of the SNe given the year it was observed. The efficiency correction ranges between 1.4 and 2.6, with a mean weighting of 1.9 for each host galaxy. By dividing the number of host galaxies by the corresponding number of field galaxies, and including a correction for the survey's observing period, we can determine how the rate of SNe Ia varies as a function of the stellar mass of their host galaxy. Figure~\ref{fig:massrate} shows the SN Ia rate for both the passive and star-forming galaxy samples. It is clear that the rate of SNe in all types of galaxies depends on the stellar mass. We also see that the relationship between the SN Ia rate and stellar mass is different for passive galaxies and star-forming galaxies in the SDSS data. 

%%%%%%%%%%%%%%%%%%%%%%%%%%  Figure 3  %%%%%%%%%%%%%%%%%%%%%%%%%%%%%%
%%%%%%%%%%%%%%%%%%%%%%%%%%%%%%%%%%%%%%%%%%%%%%%%%%%%%%%%%%%%%

The data are fit to a linear function in log-space, corresponding to a power law dependence of SN rate on stellar mass as shown on Figure~\ref{fig:massrate}. A linear dependence on stellar mass would result in a slope of unity. The error bars shown and fitting errors include statistical errors only. The uncertainty in galaxy stellar mass is discussed in \S\ref{subsec:montecarlo}. For passive galaxies, we find a best-fit slope of $0.67 \pm 0.15$ (with  ${\chi}^{2}$-statistic $({\chi}^{2}) = 2.30$ for 6 degrees of freedom) compared to a value of $0.94 \pm 0.08$ (${\chi}^{2} = 9.28$ for 6 degrees of freedom, denoted as $n_{\textrm{star-forming}}$) for star-forming galaxies. The value for passive galaxies is incompatible (at the $2.2\sigma$ level) with a linear relationship, as favored by \citet{2006ApJ...648..868S}, who found a slope of $1.10 \pm 0.12$ using the SNLS data at higher redshift. 

Figure~\ref{fig:massrate} also shows the results for passive galaxies from \citet{2006ApJ...648..868S} as open circles. We see that the SDSS galaxy sample contains fewer SNe Ia in high mass passive galaxies than SNLS and more SNe Ia in low mass passive systems.  While the two analyses should be directly comparable, the galaxy population is expected to evolve between $z \sim 0.75$ and $z \sim 0.2$.  However, it does not seem that galaxy evolution can explain these differences as more massive galaxies should be found in the local universe. In addition, we note that the SDSS analysis finds a larger slope for star-forming galaxies compared to passives, while the opposite is seen in the SNLS data, who find values of $n_{\textrm{star-forming}}=0.66 \pm 0.08$ and $0.74 \pm 0.08$, for moderately and highly star-forming galaxies respectively. \citet{2010arXiv1006.4613L} find the slope is independent of host galaxy type, with a value of $0.5$ providing a good fit in all cases.  

\subsection{Parameterizing the SN Ia Rate} 
\label{sec:AplusB}

The data in Figure~\ref{fig:massrate} indicate that the SN Ia rate depends on galaxy stellar mass, but also that the rate depends on whether the galaxy is actively forming stars.  A two-component model was considered by \citet{2006MNRAS.370..773M}, and \citet{2005ApJ...629L..85S}, who modelled the SN Ia rate of a galaxy to consist of a ``delayed" component, with a long delay time that is driven by the stellar mass of the galaxy, and a ``prompt" component, with short delay times that is caused by the formation of new stars. The model assumes that the ``delayed" component is proportional to the stellar mass independent of the galaxy age and star-formation history, and that the ``prompt" component time scale is short compared to changes in the star-formation rate.  These assumptions result in an expression whose parameters can be determined from data as was done by \citet{2006ApJ...648..868S}. In detail, the $\snrIa$ can be written as;
\begin{equation}
\label{eq:basic}
\snrIa(t) = A \times {M}(t) + B \times \dot{M}(t),
\end{equation}
where $\snrIa(t)$ is the explosion rate of SNe Ia at time $t$, ${M}(t)$ is the stellar mass of a galaxy, $\dot{M}(t)$ is the rate of change of stellar mass, and  $A$ and $B$ are constants determined from the data and have units \Aunits and \Bunits, respectively.  We assume $\dot{M}(t)$ is equal to the star-forming rate (SFR) (averaged over the previous 0.5 Gyr) as discussed in \S\ref{subsec:galprop}.   While the model is, in principle, valid for all $t$, our SN rate measurements apply only to the current era and we will suppress the dependence on $t$.  This model is commonly known as the ``A+B" model for the supernova rate and assumes that the SN Ia rate is linearly dependent on both the stellar mass of a galaxy and its star-formation rate. However, in \S\ref{sec:massrate} we showed that for passive galaxies (whose SN Ia rate will be purely dependent on stellar mass in this parameterization), a linear dependence was not favored by the SDSS dataset. We therefore generalize Equation~\ref{eq:basic} to,
\begin{equation}
\label{eq:complex}
\snrIa = A \times M^{\nmass} + B \times \dot{M}^{\nsfr},
\end{equation}
where \nmass, \nsfr, $A$ and $B$ are constants to be determined from the data. Since passive galaxies have $\dot{M}=0$, we can apply the results of \S\ref{sec:massrate} to conclude $\nmass = 0.67 \pm 0.15$.  The straight line fit to the passive galaxies yields $\log A = -9.95 \pm 0.68$ or $A = {1.11}^{+4.17}_{-0.88} \times {10}^{-10} \Aunits$. If we assume $\nmass \equiv 1$, we find a value of $\log A = -13.56 \pm 0.08$ or $A = {2.75}^{+0.57}_{-0.47} \times {10}^{-14}$, which differs at $2.1\sigma$ with the value of $A = 5.3 \pm 1.1 \times {10}^{-14}$ found using the SNLS dataset.

While the above parameterization of the SN Ia rate uses the stellar mass and the recent star-formation rate, other galaxy properties can be considered, such as the metallicity, age and level of extinction. \citet{2005ApJ...634..210G} find qualitative evidence suggesting that the progenitor age is a possible source of diversity in SNe Ia properties. However, there is a degeneracy between the age of a galaxy, and its metallicity, which is extremely difficult to break using broad-band photometry. We thus confine ourselves to considering the stellar mass and star-formation rates of our host galaxies in this analysis, but note that with improved stellar population models, a larger wavelength coverage and galaxy spectra, it may be possible to break this degeneracy. Using SDSS-II SNe, \citet{Gupta2011} attempt to break this degeneracy by using multi-wavelength photometry to better constrain the ages of their SN Ia host galaxies while \citet{DAndrea2011} and \citet{Konishi2011} use spectral features to determine the metallicities of their host galaxies.

\subsection{SN Ia Rate as a Function of Host Galaxy Mean Star-formation Rate}
\label{sec:sfrrate}

\begin{figure*}
\epsscale{1.5}
\plotone{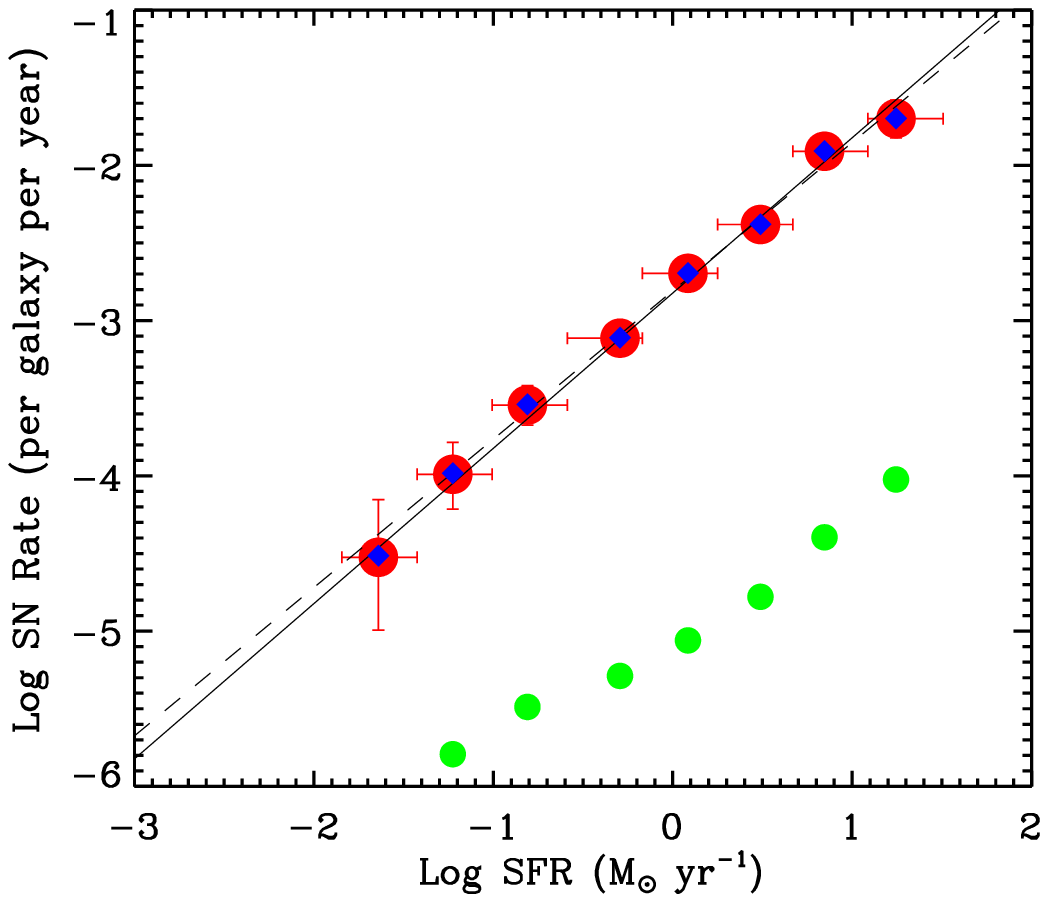}
\plotone{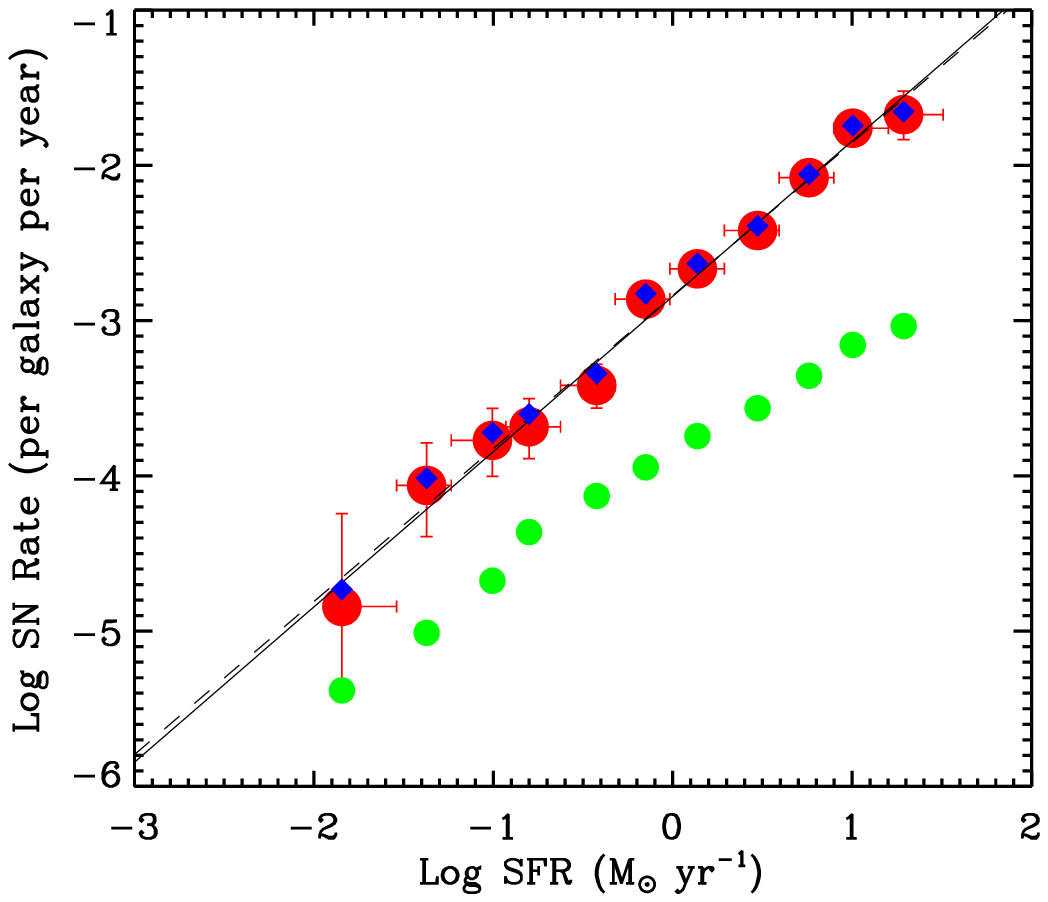}
\caption{SN Ia rate as a function of host galaxy star-formation rate. \emph{Left panel:} Green points indicate the expected rate of SNe Ia due to the stellar mass of each galaxy, using the values of $\nmass$ and $A$ as determined in \S\ref{sec:massrate}. Blue diamonds show the observed rate of SNe Ia per galaxy per year, while the red points are the excess (\textit{i.e.} the difference between the blue and green values). A best-fitting line (dashed), and best-fitting line with unit slope (solid) is also shown. \emph{Right panel:} Identical, except a value of $\nmass \equiv 1$ is assumed. \label{fig:sfrrate}}
\end{figure*}

We now consider the star-forming galaxies to determine $B$ and $\nsfr$.  We bin the host galaxy and comparison field sample in star-formation rate, and as in \S\ref{sec:massrate}, correct both samples for incompleteness, using the SN efficiency for the host galaxy sample and the $1/V_{\textrm{max}}$ correction for both the host galaxy and comparison field samples. The SN Ia rate is shown (blue diamonds) as a function of SFR in Figure~\ref{fig:sfrrate}.  We want to determine the excess SN Ia rate due to recent star-formation activity assuming that the term proportional to stellar mass is the same for star-forming and passive galaxies. The portion due to the stellar mass term is calculated using Equation~\ref{eq:complex}, and shown on the figures (green points) as are the SN Ia rates after the stellar mass term has been subtracted (red points). The left panel of Figure~\ref{fig:sfrrate} uses our best-fit line with slope $\nmass = 0.67$ while the right panel uses the fit where the slope is fixed at  $\nmass \equiv 1$. 

The observed SN Ia rate depends strongly on recent star-formation and greatly exceeds the rate in passive galaxies with identical stellar mass. We find  $\nsfr=0.96\pm0.07$ and $\log B = -2.81\pm0.04$ ($B={1.55}^{+0.16}_{-0.15} \times {10}^{-3} \Bunits$) with ${\chi}^{2} = 1.58$ for 6 degrees of freedom when  $\nmass = 0.67$. When $\nmass \equiv 1$ is assumed, we find $\nsfr =  0.98 \pm 0.08$, and $\log B = -2.85\pm0.05$ ($B={1.42}^{+0.17}_{-0.15} \times {10}^{-3} \Bunits$) with ${\chi}^{2} = 1.52$ for 6 degrees of freedom. The lack of sensitivity to the value of $\nmass = 0.674$ follows because the stellar mass term is always small compared to the star-forming term.  
 
Our best-fit to Equation~\ref{eq:complex}, is therefore
%%%%%%%%%%%%%%%%%%%%%%%%%%  Figure 4  %%%%%%%%%%%%%%%%%%%%%%%%%%%%%%
%%%%%%%%%%%%%%%%%%%%%%%%%%%%%%%%%%%%%%%%%%%%%%%%%%%%%%%%%%%%%%

\begin{equation}
\label{eq:best}
\begin{split}
\textrm{SNR}_{\textrm{Ia}} = {1.11}^{+4.17}_{-0.88} \times {10}^{-10} \mgal^{0.67 \pm 0.15} \\
+ {1.55}^{+0.16}_{-0.15} \times {10}^{-3} {\dot{M}}^{0.96 \pm 0.07}.
\end{split}
\end{equation}

As noted previously, the analysis of \citet{2006ApJ...648..868S} in the redshift range $0.2 < z < 0.75$ preferred a SN Ia rate linearly dependent to the stellar mass of a galaxy. If we assume $\nmass \equiv 1$ and $\nsfr \equiv 1$, we find,
\begin{equation}
\label{eq:nmass1nsfr1}
\snrIa = {2.75}^{+0.57}_{-0.47} \times {10}^{-14} \mgal + {1.40}^{+0.14}_{-0.13} \times {10}^{-3} \dot{M}.
\end{equation}

For comparison, \citet{2006ApJ...648..868S} find values of $A = 5.3 \pm 1.1 \times {10}^{-14} \Aunits$ and $B = 3.9 \pm 0.7 \times {10}^{-4} \Bunits$. Our value of $A$ is $2.1 \sigma$ lower, while the values of $B$ are inconsistent at $3.5 \sigma$, indicating that recent star-formation activity plays a more significant role in determining the overall SNe Ia rate for our sample. This result is consistent with models of how galaxies evolve through cosmic time. Observations suggest, that at high redshift ($z=0.75$), the rate of star-formation is far higher than in the local Universe. Combining this with measurements suggesting that the SN Ia rate increases slowly as a function of redshift, suggests that recent star-formation activity is more significant in determining the SN Ia rate at low redshift. The methodology used in this analysis is similar to that used in \citet{2006ApJ...648..868S}, and has been significantly tested (\S\ref{app:color-mag}, \S\ref{app:pegase_photoz}, \S\ref{app:effect}, \S\ref{app:restwave}). 

\subsection{Bivariate Fitting}
\label{sec:bivariate}

Thus far we have used only the passive galaxies to determine the $A$ term and then used the star-forming galaxies to determine the $B$ term, while keeping $A$ fixed. A more sophisticated method is to constrain the parameters simultaneously using all galaxy types, thus making optimal use of the data. We bin the host galaxy and comparison field sample in the stellar mass and star-formation plane, and correct for incompleteness. By dividing the number of host galaxies in each bin by the corresponding number of field galaxies, we are able to determine the SN Ia rate in each bin of stellar mass and star-formation rate. We consider several variations on Equation~\ref{eq:complex}. First, we consider the case where $B \equiv 0$, \textit{i.e.} the SN Ia rate is purely dependent on stellar mass, and $\nmass$ is a free parameter. In this case, we find $A=1.08 \pm 0.18 \times {10}^{-10}$ and $\nmass = 0.68 \pm 0.005$, in agreement with the result found in \S\ref{sec:massrate}. However, this is a poor fit to the data, and allowing $B\ne0$ but assuming  $\nsfr \equiv 1$ reduces the $\chi^2$ from $347$ for $42$ degrees of freedom to $142$ for $41$ degrees of freedom, and yields values of $\nmass$, $A$, and $B$ consistent with those found in \S\ref{sec:sfrrate}. Finally, we allow $\nsfr$ to vary and find $\nsfr =  1.00 \pm 0.05$ with $\chi^2=142$ for $40$ degrees of freedom, a negligible improvement. 

We thus conclude that our data is consistent with a linear dependence on star-formation rate. Our fiducial result using bivariate fitting is
\begin{equation}
\label{eq:bestfit_linearsfr}
\begin{split}
\textrm{SNR}_{\textrm{Ia}} = 1.05 \pm 0.16 \times {10}^{-10} \mgal^{0.68 \pm 0.01} \\
 + 1.01 \pm 0.09 \times {10}^{-3} {\dot{M}^{1.00 \pm 0.05}}.
\end{split}
\end{equation}
This is in good agreement with the values found in \S\ref{sec:massrate} and \S\ref{sec:sfrrate}. 

\subsection{SN Ia Rate as a Function of Specific Star-Formation Rate}
\label{sec:ssfr}

The results from \S\ref{sec:massrate}, \S\ref{sec:sfrrate} and \S\ref{sec:bivariate} have shown that the SN Ia rate depends on both the galaxy stellar mass and star-formation rate, with star-formation rate dominating the SN Ia rate. Here, we study how the SN Ia rate is related to host galaxy type. To determine this, we bin the host galaxy and comparison field samples according to their value of sSFR. Both samples are corrected for incompleteness, and the total stellar mass of the field sample is calculated. By dividing the incompleteness corrected number of host galaxies by the total stellar mass of the field sample, we are able to determine the SN Ia rate per unit stellar mass as a function of sSFR. As noted in \S\ref{sec:hostgalprop}, sSFR is a way of distinguishing between galaxy types, with galaxies with low values of sSFR being primarily large galaxies that are using a small fraction of their total stellar mass to form new stars, while those with larger levels of sSFR are starburst galaxies, or galaxies that are using a significant fraction of their stellar mass to form new stellar systems. 

%%%%%%%%%%%%%%%%%%%%%%%%%%  Figure 5  %%%%%%%%%%%%%%%%%%%%%%%%%%%%%%
\begin{figure*}[ht]
\epsscale{1.5}
\plotone{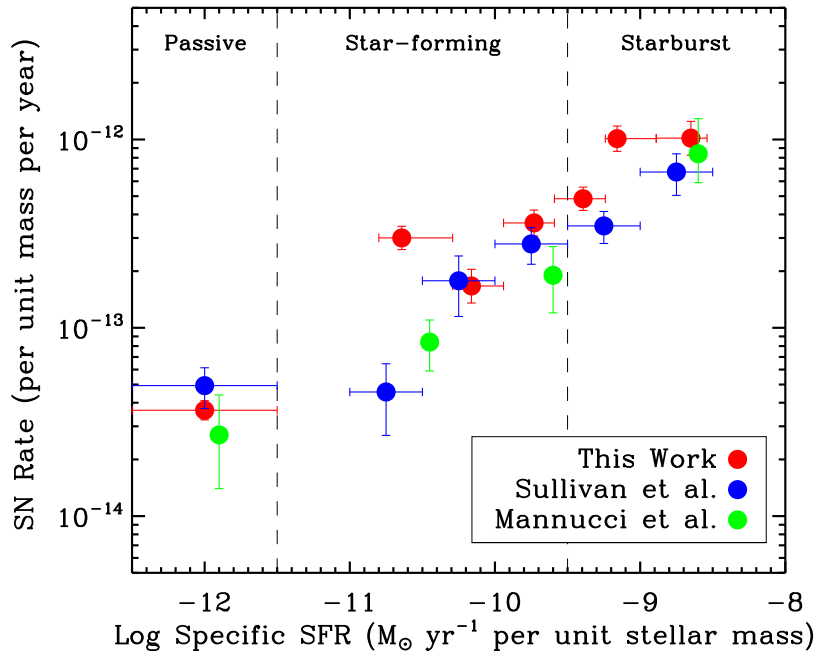}
\caption{The SN Ia rate per unit stellar mass per year as a function of host galaxy specific star-formation rate (sSFR). The red points are those determined by the SDSS analysis, blue points are those from \citet{2006ApJ...648..868S}, while points shown in green are measurements at low redshift made by \citet{2005A&A...433..807M}, where the magnitude and color of the host galaxy have been used to determine the host galaxy stellar mass and star-formation rate. The horizontal errors on the SDSS data indicate the bin width while the horizontal positions represents the mean of the data in that bin. The positioning of green points on the x-axis is somewhat uncertain since precise values for sSFR were not given. Passive galaxies have $\mathrm{sSFR}=0$, but are shown on this graph, with $\log \mathrm{sSFR} \simeq -12$. }\label{fig:ssfr}
\end{figure*}
%%%%%%%%%%%%%%%%%%%%%%%%%%%%%%%%%%%%%%%%%%%%%%%%%%%%%%%%%%%%%

Figure~\ref{fig:ssfr} exhibits the rate of SNe Ia per unit stellar mass in star-forming galaxies as a function of sSFR.  The rate increases with sSFR, reaching an increased factor of $\sim 30$ for starburst galaxies compared to passive galaxies. The measurements of this work are in excellent agreement with those found at higher redshift \citep{2006ApJ...648..868S} and in the local universe \citep{2005A&A...433..807M}, indicating that this relationship holds for all redshifts that have been studied. The SDSS data, however, has a point that appears to disagree with the other data and the generally linear trend of increasing SN Ia rate with sSFR.  This point corresponds to the galaxies highlighted in \S\ref{sec:hostgalprop} as being on the edge of the moderately star-forming galaxies. Appendix \ref{app:color-mag} considers these objects, and determines that while there was a possible ambiguity in the classification of these objects, they constitute a distinct population that lie between passive and star-forming galaxies.  Any contamination by passive galaxies would tend to reduce the rate, but the measurement appears to be high compared to previous measurements.  Another possibility is that we might have underestimated the number of weakly star-forming field galaxies but there is no evidence that this is the case based on the comparison in Appendix \ref{app:brinc}.

\subsection{The Effect of our Selection Criteria}
\label{sec:selectioncuts}

We have studied how the SN Ia rate is related to the host galaxy properties for the SDSS sample. However, as discussed in \S\ref{sec:spectroissues}, the sample constructed for this analysis is comprised of SNe Ia that have not all been spectroscopically confirmed and thus may be contaminated by non-SN Ia events. Our analysis has also used an efficiency correction, which is increasingly important towards the edge of our redshift range, and thus can cause uncertainties in our results. 

Table~\ref{tab:hostsystematics} shows the results that we obtain using various subsets of our SN Ia sample. In the two left-hand columns we show fits for the spectroscopically confirmed and unconfirmed portions of our sample.  In the three right-most columns we show the results for three different redshift ranges.  The spectroscopically confirmed and unconfirmed subsamples are, of course, incomplete, so the $A$ and $B$ parameters will necessarily be smaller than for the full sample.  The results for $\nmass$ and $\nsfr$, however, should be comparable. 

From Table~\ref{tab:hostsystematics} we see that the spectroscopically confirmed sample is fit by $\nmass=0.873 \pm 0.273$, consistent with the combined result but also consistent with $\nmass=1$. This value of $\nmass$ may be due to the lower proportion of passive galaxies in this sample and a bias against more luminous and thus massive galaxies in the spectroscopic selection. This bias is caused by a targeting against probable SNe Ia that occur in the centres of luminous galaxies, making them difficult to identify spectroscopically. As the redshift limit considered is decreased, resulting in a more complete sample, the value of $\nmass$ is stable and shows no trend towards one (although the errors increase rapidly as the sample size is reduced). The value for $\log A$ when we assume $\nmass \equiv 1$ is consistent for all the redshift ranges (when $\nmass$ is a free parameter, it is highly degenerate with $\log A$). Table~\ref{tab:hostsystematics} also shows that the value of $\nsfr$ is not influenced by the inclusion of non-spectroscopically confirmed SNe Ia nor the redshift range. 

Table~\ref{tab:hostsystematics} also shows how our selection criteria affects the dependence that the SN Ia rate has on the star-formation rate. We showed in \S\ref{sec:sfrrate} and \S\ref{sec:bivariate}, that the SN Ia rate depends approximately linearly on the recent star-formation rate. The subsamples displayed in Table~\ref{tab:hostsystematics} are all consistent and there is no hint on any deviation from $\nsfr \sim 1$. A value of $\log B \sim -2.85$ is valid for all redshift ranges considered. 

Finally, we study the results of \S\ref{sec:ssfr}. The ``passive rate" in Table~\ref{tab:hostsystematics} is the rate per unit stellar mass per year in passive galaxies, as shown in Figure~\ref{fig:ssfr}, while the ``starburst rate is the corresponding rate for galaxies with the highest levels of sSFR ($\textrm{sSFR} > -9.5$). The values of both the passive and starburst rates are consistent independent of the redshift limit used and are consistent with the sum of the spectroscopically confirmed and unconfirmed samples.  In all cases, the rate of SNe Ia per unit stellar mass in highly star-forming galaxies is significantly higher (by a factor of $\sim 30$) than that seen in passive galaxies. 

\begin{deluxetable}{ccccccc}
\tabletypesize{\scriptsize}
\tablecaption{Effect of our selection criteria on the results described in \S\ref{sec:massrate}, \S\ref{sec:sfrrate} and \S\ref{sec:ssfr} \label{tab:hostsystematics}}
\tablewidth{0pt}
\tablehead{ \colhead{Parameter} & \colhead{Nominal Result} & \colhead{Confirmed$^g$} & \colhead{Phot-ID$^h$} & \colhead{$z<0.20$} & \colhead{$z<0.16$} & \colhead{$z<0.12$}}
\startdata
No. Hosts & $342$ & $197$ & $145$ & $196$ & $103$ & $36$ \\
\% Confirmed SNe & $57.6$ & $100.0$ & $0.0$ & $67.9$ & $79.6$ & $97.2$ \\
$\nmass$ & $0.67 \pm 0.15$ & $0.87 \pm 0.27$ & $0.43 \pm 0.41$ & $0.66 \pm 0.20$ & $0.62 \pm 0.26$ & $0.62 \pm 0.93$ \\		
$n_{\textrm{star-forming}}$ & $0.94 \pm 0.08$ & $0.93 \pm 0.11$ & $0.92 \pm 0.12$ & $0.82 \pm 0.09$ & $0.82 \pm 0.15$ & $0.77 \pm 0.28$ \\		
$\log A$ $^a$ & $-9.95 \pm 0.68$ & $-12.42 \pm 0.89$ & $-7.74 \pm 1.04$ & $-10.00 \pm 0.76$ & $-9.47 \pm 0.86$ & $-9.30 \pm 1.37$ \\	
$\log A$ $^b$ & $-13.56 \pm 0.08$ & $-13.77 \pm 0.11$ & $-13.88 \pm 0.13$ & $-13.68 \pm 0.11$ & $-13.61 \pm 0.14$ & $-13.38 \pm 0.24$ \\	
$\nsfr$ $^a$ & $0.96 \pm 0.07$ & $0.95 \pm 0.11$ & $0.96 \pm 0.15$ & $1.00 \pm 0.10$ & $1.01 \pm 0.12$ & $1.14 \pm 0.29$ \\	
$\nsfr$ $^b$ & $0.98 \pm 0.08$ & $1.02 \pm 0.11$ & $0.98 \pm 0.15$ & $1.08 \pm 0.12$ & $1.07 \pm 0.16$ & $1.13 \pm 0.47$ \\	
$\log B$ $^c$ & $-2.81 \pm 0.04$& $-3.06 \pm 0.06$ & $-3.18 \pm 0.07$ & $-2.88 \pm 0.06 $ & $-2.91 \pm 0.08$ & $-2.90 \pm 0.15$ \\
$\log B$ $^d$ & $-2.81 \pm 0.04$ & $-3.08 \pm 0.05$ & $-3.19 \pm 0.07$ & $-2.88 \pm 0.05$ & $-2.91 \pm 0.08$ & $-2.93 \pm 0.15$ \\
$\log B$ $^e$ & $-2.85 \pm 0.04$ & $-3.09 \pm 0.06$ & $-3.21 \pm 0.07$ & $-2.91 \pm 0.06$ & $-3.01 \pm 0.10$ & $-3.05 \pm 0.18$ \\
\% Passive galaxies & $23.4$ & $20.3$ & $27.6$ & $25.0$ & $28.2$ & $33.3$\\
Passive Rate $^f$ & $3.6 \pm 0.6$ & $1.7 \pm 0.4$ & $1.9 \pm 0.5$ & $2.8 \pm 0.6$ & $3.2 \pm 0.9$ & $3.8 \pm 1.8$\\	
Starburst Rate $^f$& $124.7 \pm 29.95$ & $66.3 \pm 24.2$ & $35.3 \pm 19.8$ & $119.4 \pm 56.7$ & $96.0 \pm 68.9$ & $117.9 \pm 109.0$ \\
\enddata
\tablenotetext{a}{$\nmass$ free, in units of $\Aunits$}
\tablenotetext{b}{$\nmass \equiv 1$, in units of $\Aunits$}
\tablenotetext{c}{$\nmass$ and $\nsfr$ free, $\Bunits$}
\tablenotetext{d}{$\nmass$ free and $\nsfr \equiv 1$, $\Bunits$}
\tablenotetext{e}{$\nmass \equiv 1$ and $\nsfr \equiv 1$, $\Bunits$}
\tablenotetext{f}{$\times {10}^{-14}$ per unit mass per year}
\tablenotetext{g}{Considering solely spectroscopically confirmed SN Ia}
\tablenotetext{h}{Considering solely photometrically typed SN Ia}
\end{deluxetable}

The fit parameters shown in Table~\ref{tab:hostsystematics} do not evolve across our redshift range. Since any evolution would be unexpected due to the small range in cosmic time covered by our analysis, it is reassuring to note that our results are insensitive to the redshift interval that is chosen.  The only parameter that significantly changes with redshift is the proportion of passive galaxies found. This may simply reflect observations that SNe Ia in passive galaxies are fainter than their star-forming counterparts (see \S\ref{sec:sneprop} for an analysis with the SDSS sample) and thus are not observed at higher redshifts by the SDSS-II SN survey. 

In Table~\ref{tab:hostsystematics} we considered separately the spectroscopically confirmed and unconfirmed SNe Ia and various redshift ranges on our conclusions. However, there are several other uncertainties that can arise as part of our selection criteria and analysis. We also investigated the effect of using different priors on $A_V$ when determining the sample of SNe Ia, by using a flat prior and a positive prior ($A_V\ge0$). The various priors produce results that are entirely consistent with those found previously. In the determination of $\nmass$ and $\nsfr$ we have performed linear fits to the log-log plots, but it is also possible to fit to the power law form directly. These fits are consistent with our linear fits to the logarithms. We have also considered various bin sizes for each stage of our analysis and find that our results are unaffected. We also considered the possibility that our results may depend on a specific, anomalous year with the SDSS-II SN survey or may vary as a function of position on the sky. We split the host galaxy sample by both year and position but found no variation on our final results. Finally, we considered the effect of modelling $\epsilon$, the survey efficiency, as a function of redshift, $A_V$ and $\Delta$, to account for the observation that passive galaxies host fainter, higher $\Delta$, SNe than star-forming galaxies (\S\ref{sec:sneprop}), which could result in the survey efficiency varying as a function of host galaxy type. This additional correction, which has not been applied for other previous analyses, produces results that are entirely consistent with our fiducial result.

In order to study the robustness of our results, we have considered the effect of altering our selection criteria, in Table~\ref{tab:hostsystematics}. We have shown that the inclusion of non-spectroscopically confirmed SNe Ia in our sample and varying our redshift range considered does not significantly change the values of $A, B, \nmass$ and $\nsfr$

\subsection{The Effect of SED Errors on our results}
\label{subsec:montecarlo}

For each host galaxy in our sample we have determined a value for its stellar mass and recent star-formation rate. Each of these measurements has an associated error that may allow galaxies to move between bins, and thus affect our fitted parameters. This may be especially important for the sample of galaxies with ambiguous classification, highlighted in \S\ref{sec:hostgalprop} and Appendix \ref{app:color-mag}.  

To quantify this effect on our results we use both a Monte-Carlo (MC) and Bootstrap (BP) approach. For the MC analysis 10,000 realizations of the host galaxy sample are made by drawing from the estimated probability distribution for each host. We consider two cases: varying the stellar mass of each host galaxy and varying the stellar mass and star-formation rate. The second case allows galaxies to move from passive to star-forming and vice versa. Thus each of the MC samples consists of 342 host galaxies, each a variation on one specific host in the host galaxy sample. 

For the BP analysis, we again obtain 10,000 realizations of the host galaxy sample, this time by selecting a host galaxy at random with replacement. This analysis allows each host galaxy to be selected on multiple occasions, probing the effect that outliers within the sample may have on our results. As with the MC approach we consider two cases: selecting the host galaxies before the sample has been separated into passive and star-forming datasets (thus allowing the relative proportions to change) and randomly sampling after separation, thus enforcing the same proportion of passive and star-forming galaxies in each dataset. The second approach tests the dependence of our results on a subset of objects, while the first case is particularly important for the sample of ambiguous galaxies, identified in \S\ref{sec:hostgalprop}, and investigates if they are likely to be predominantly passive in nature. 

To determine how the SED uncertainties affect our overall conclusions, we determine the value of each parameter for each realization, and fit a Gaussian (which is observed to provide a good fit) to each distribution. This provides an estimate for the central values and systematic uncertainty in each case. Table \ref{tab:montecarlo} gives the values and associated errors for the parameters determined in this work for each of these four systematic tests. 

\begin{deluxetable}{ccccccc}
\tabletypesize{\scriptsize}
\tablecaption{Effect of our SED uncertainty on the results described in \S\ref{sec:massrate}, \S\ref{sec:sfrrate} and \S\ref{sec:ssfr} \label{tab:montecarlo}}
\tablewidth{0pt}
\tablehead{ \colhead{Parameter} & \colhead{Nominal Result} & \colhead{MC (variable M)} & \colhead{MC (variable M and SFR)} & \colhead{BP (split)} & \colhead{BP (not split)}}
\startdata
$\nmass$ & $0.67 \pm 0.15$ & $0.76 \pm 0.07$ & $0.78 \pm 0.08$ & $0.68 \pm 0.11$ & $0.68 \pm 0.11$ \\		
$n_{\textrm{star-forming}}$ & $0.94 \pm 0.08$ & $0.93 \pm 0.03$ & $0.93 \pm 0.04$ & $0.94 \pm 0.07$ & $0.94 \pm 0.07$ \\
$\log A$ $^a$ & $-9.95 \pm 0.68$ & $-10.98 \pm 0.73$ & $-11.14 \pm 0.82$ & $-10.14 \pm 1.13$ & $-10.15 \pm 1.12$ \\	
$\log A$ $^b$ & $-13.56 \pm 0.08$ & $-13.49 \pm 0.04$ & $-13.51\pm 0.04$ & $-13.53 \pm 0.08$ & $-13.54 \pm 0.09$ \\	
$\nsfr$ $^a$ & $0.96 \pm 0.07$ & $0.96 \pm 0.001$ & $0.85 \pm 0.10$ & $0.97 \pm 0.06$ & $0.97 \pm 0.06$ \\	
$\nsfr$ $^b$ & $0.98 \pm 0.08$ & $0.99 \pm 0.001$ & $0.85 \pm 0.09$ & $1.00 \pm 0.06$ & $1.00 \pm 0.06$ \\	
$\log B$ $^c$ & $-2.81 \pm 0.04$ & $-2.81 \pm 0.001$ & $-2.85 \pm 0.04$ & $-2.81 \pm 0.03$ & $-2.81 \pm 0.03$ \\
$\log B$ $^d$ & $-2.81 \pm 0.04$ & $-2.83 \pm 0.001$ & $-2.90 \pm 0.07$ & $-2.82 \pm 0.03$ & $-2.82 \pm 0.03$ \\
$\log B$ $^e$ & $-2.85 \pm 0.04$ & $-2.85 \pm 0.001$ & $-2.94 \pm 0.09$ & $-2.86 \pm 0.02$ & $-2.86 \pm 0.02$ \\
Passive Rate $^f$ & $3.6 \pm 0.6$ & $3.6 \pm 0.01$ & $3.3 \pm 0.1$ & $3.6 \pm 0.4$ & $3.55 \pm 0.4$ \\	
Starburst Rate $^f$& $124.7 \pm 29.95$ & $92.2 \pm 11.4$ & $85.4 \pm 13.8$ & $102.4 \pm 18.8$ & $102.8 \pm 19.0$ \\
\enddata
\tablenotetext{a}{$\nmass$ free, in units of $\Aunits$}
\tablenotetext{b}{$\nmass \equiv 1$, in units of $\Aunits$}
\tablenotetext{c}{$\nmass$ and $\nsfr$ free, \Bunits}
\tablenotetext{d}{$\nmass$ free and $\nsfr \equiv 1$, \Bunits}
\tablenotetext{e}{$\nmass \equiv 1$ and $\nsfr \equiv 1$, \Bunits}
\tablenotetext{f}{$\times {10}^{-14}$ per unit mass per year}
\end{deluxetable}

Both the MC and BP analysis provide values for the parameters determined that are consistent with those found as our main result, as described in \S\ref{sec:massrate}, \S\ref{sec:sfrrate}, \S\ref{sec:ssfr}. In all cases considered the observed scatter from the MC and BP tests is smaller than the statistical uncertainty. We note that while the central value for $\nmass$ determined by the MC analysis is larger than our default result, it is still inconsistent with $\nmass=1$ at the $3.4\sigma$ level when the stellar mass is allowed to vary, and $2.9 \sigma$ when galaxies are allowed to move from passive to star-forming. 

We observe that the SN Ia rate per unit stellar mass in passive galaxies is consistent in all four cases considered. This implies that the sample of galaxies with ambiguous classifications (as noted in \S\ref{sec:hostgalprop}) do not affect our overall conclusions. In the MC where the stellar mass and star-formation rate are allowed to vary, these galaxies are able to move from moderately star-forming to passive where their error bars allow. However, we note that in this case, the SN Ia rate in passive galaxies is in fact lower than the observed value, suggesting that these galaxies are not passively evolving, and may have non-zero star-formation rates. 

The systematic error bars determined by the MC and BP tests are sub-dominant to the statistical uncertainties obtained in \S\ref{sec:massrate}, \S\ref{sec:sfrrate} and \S\ref{sec:ssfr}. Therefore, the uncertainties due to the SED fitting are not the major source of uncertainty. Extensive testing of the \PEGASE SEDs is carried out in Appendix~\ref{app:color-mag},~\ref{app:pegase_photoz},~\ref{app:brinc} and~\ref{app:restwave}. An offset is found in the photometric redshift estimates and associated stellar mass estimates for the comparison field sample used in our analysis. By assuming that this offset does not affect the star-formation rates for our galaxies, which are inferred through their color (which we determine in Appendix~\ref{app:color-mag}), we uniformly apply this offset to each galaxy in the comparison field sample, and recalculate the results of \S\ref{sec:massrate}, \S\ref{sec:sfrrate} and \S\ref{sec:ssfr}. While we find some variation in the central values, our conclusions are unaffected. We find that the rate of SNe Ia as a function of stellar mass in passive galaxies is incompatible with a linear relationship in all cases considered. Having applied the determined offset we find a value of $\nmass \sim 0.5$ is preferred, which is consistent with our result at the $1.2\sigma$ level. The excess rate of SNe Ia in star-forming galaxies is linearly proportional to the star-formation rate in all cases considered. We find some evidence for a lower SNe Ia rate per unit stellar mass in passive galaxies, than determined in \S\ref{sec:ssfr}, with a maximum difference of $2.2\sigma$. 

\subsection{Comparison to other results}
\label{sec:hostcomp}

Throughout this analysis we have compared our results to that of \citet{2006ApJ...648..868S}. We find a different dependence on stellar mass for the SN Ia rate, but agree that there is a strong dependence on the recent star-formation rate. In agreement with the results of \citet{2006ApJ...648..868S}, we find that the SN Ia rate per unit stellar mass is greater in highly star-forming galaxies compared to passive galaxies, with an approximately linear dependence on sSFR except for one potentially anomalous point in the SDSS data. Different assumptions about the dependence of the SN Ia rate on stellar mass do not significantly alter our conclusions about its dependence on the recent star-formation rate. 

A summary of how our results compare to those found by other studies is given in Table~\ref{tab:hostcomparison}. \citet{2005ApJ...629L..85S,2005A&A...433..807M} also investigated the possibility that the SN Ia rate may be a two-component model, assuming $\nmass=\nsfr=1$. These analyses updated SN rates from \citet{1999A&A...351..459C}, determining values of $A = {3.83}^{+1.4}_{-1.2} \times {10}^{-14} \Aunits$, for SNe in E/S0 galaxies (which can be crudely associated with passive galaxies in this analysis) and either $B = {1}^{+0.6}_{-0.5} \times {10}^{-3} \Bunits$ or $B = {2.3} \pm 1. \times {10}^{-3} \Bunits$, depending on whether the $z \le 1.0$ core-collapse SN rate density \citep{2004ApJ...613..189D} compared to the star-formation rate density \citep{2004ApJ...600L.103G}, or the population of SNe Ia found in blue ($B-K$) galaxies is used to model the SN Ia rate in star-forming galaxies. The value of $A$ is consistent with our result (${2.8}^{+0.6}_{-0.5} \times {10}^{-14} \Aunits$), when $\nmass \equiv 1$ is assumed, and our value of $B = {1.4}^{+0.2}_{-0.1} \times {10}^{-3} \Bunits$ when $\nmass \equiv 1$ and $\nsfr \equiv 1$ is also in good agreement. However, as noted, the A+B model for the SN Ia rate that depends linearly on $M$ and $\dot{M}$ does not provide the best-fit for our dataset. \citet{2005ApJ...629L..85S,2005A&A...433..807M} assumed that the SN Ia rate depends linearly on  $M$ and $\dot{M}$. \citet{2008ApJ...682..262D} using a sample of low redshift SNe ($z<0.12$) from the SDSS-II SN Survey (and overlapping with this work) combined with other published work, used the global star-formation rate as determined by \citet{2006ApJ...651..142H} (which may over-estimate the total mass density) to determine $A = (2.8 \pm 1.2) \times {10}^{-14} \Aunits$ and $B = (0.93 \pm 0.34) \times {10}^{-3} \Bunits$, which are in agreement with those found in this analysis. \citet{2010arXiv1006.4613L} study how the SN Ia rate is related to the size, color and morphology of the host galaxy for a sample of local SNe. They show that the SN Ia rate is not linearly related to the stellar mass of the host galaxy, instead preferring a relationship, $\snrIa \propto {M}^{\sim 0.5}$, independent of host galaxy morphology and color. Their result for elliptical galaxies is in excellent agreement with our results for passive galaxies, favoring a SN Ia rate proportional to ${M}^{0.67 \pm 0.15}$. However, our results differ for star-forming galaxies, where we find that an $\snrIa \propto {M}^{0.94\pm0.08}$ is favored. \citet{2010arXiv1006.4613L} also consider the case where $\snrIa \propto M$ for elliptical galaxies, finding a value of $A={4.4}^{+0.9}_{-0.8} \times {10}^{-14} \Aunits$ (in our framework), which is consistent with our result. 

\begin{deluxetable}{cccccc}
\tabletypesize{\scriptsize}
\tablecaption{A comparison of the results of \S\ref{sec:massrate} and \S\ref{sec:sfrrate} from this paper with other published analyses. \label{tab:hostcomparison}}
\tablewidth{0pt}
\tablehead{ \colhead{Analysis} & \colhead{Redshift range} & \colhead{\nmass $^a$} & \colhead{\nsfr $^b$} & \colhead{A $^a$} & \colhead{B $^b$} \\
\colhead{ } & \colhead{covered} & \colhead{ } & \colhead{ } & \colhead{\Aunits} & \colhead{\Bunits}}
\startdata
This work & $0.05 < z < 0.25$ & $0.67 \pm 0.15$ & $0.96 \pm 0.07$ & ${1.11}^{+4.17}_{-0.88} \times {10}^{-10}$ & ${1.55}^{+0.16}_{-0.15} \times {10}^{-3}$ \\
This work & $0.05 < z < 0.25$ &  fixed $=1$ &  fixed $=1$ & ${2.8}^{+0.6}_{-0.5} \times {10}^{-14}$ & ${1.4}^{+0.2}_{-0.1} \times {10}^{-3}$ \\
\citet{2006ApJ...648..868S} & $0.2 < z < 0.75$ & $1.10 \pm 0.12$ & $0.84 \pm 0.06$ & - & - \\
\citet{2006ApJ...648..868S} & $0.2 < z < 0.75$ &  fixed $=1$ &  fixed $=1$ & $5.3 \pm 1.1 \times {10}^{-14}$ & $3.9 \pm 0.7 \times {10}^{-4}$ \\
Mannucci et al. (2005) $^{c,d}$ & low redshift & fixed $=1$ & fixed $=1$ & ${3.83}^{+1.4}_{-1.2} \times {10}^{-14}$ & ${2.3} \pm 1. \times {10}^{-3}$ \\
\citet{2008ApJ...682..262D} $^e$ & $z < 0.12$ & fixed $=1$ & fixed $=1$ & $2.8 \pm 1.2 \times {10}^{-14}$ &  $0.93 \pm 0.34 \times {10}^{-3} $ \\
\citet{2010arXiv1006.4613L} $^f$ & $z < 0.05$ & fixed $=1$ & - & ${4.4}^{+0.9}_{-0.8} \times {10}^{-14}$ & - \\
\enddata
\tablenotetext{a}{As derived in \S\ref{sec:massrate}}
\tablenotetext{b}{As determined in \S\ref{sec:sfrrate}}
\tablenotetext{c}{Results taken from Mannucci et. al. (2005) and \citet{2005ApJ...629L..85S}}
\tablenotetext{d}{Approximating E/S0 galaxies for passive galaxies to determine the value of $A$ and using the rate of SNe Ia in blue ($B-K$) galaxies to determine the value of B}
\tablenotetext{e}{Using the global star-formation rate as determined by \citet{2006ApJ...651..142H}}
\tablenotetext{f}{Considering Elliptical galaxies as passive galaxies}
\end{deluxetable}

%%%%%%%%%%%%%%%%%%%%%%%%%%%%%%%%%%%%%%%%%%%%%%%%%%%%%%%%%%%%%%%%%%%%%%%%%%%%%%%%%%

\section{SNe Properties}
\label{sec:sneprop}

We have studied how the rate of SNe Ia is related to the host galaxy properties and have seen that the rate of SNe is dependent the galaxy stellar mass and star-formation rate. We extend this analysis to consider how the SN Ia light-curve parameters are related to the host galaxy properties.  Previously studies by \citet{1995AJ....109....1H,2000AJ....120.1479H} and \citet{2006ApJ...648..868S}, for example, found that bright SNe Ia are preferentially seen in young stellar environments, and \citet{1996AJ....112.2391H} showed that there is a strong correlation between the light-curve decline rate and the host galaxy morphology. The homogeneity of the SDSS-II SN sample provides an ideal opportunity to determine SN Ia light-curve parameters as a function of the galaxy star-forming rate.

SNe Ia have two key observables that affect their use as cosmological probes; their light-curve decline rate / peak brightness relationship \citep{1993ApJ...413L.105P} and their color. In MLCS2k2, the relationship between the peak luminosity of a SN Ia and the shape of its light-curve, is parameterized through the $\Delta$ parameter, where smaller $\Delta$ values correspond to brighter SNe Ia. The observed color excess of SNe Ia is modelled as the level of extinction in the $V$ band, through the parameter $A_V$. 

\subsection{MLCS2k2 $\Delta$ Parameter as a Function of Host Galaxy Type}
\label{sec:snprop_delta}

\begin{figure*}[ht]
\epsscale{1.5}
\plotone{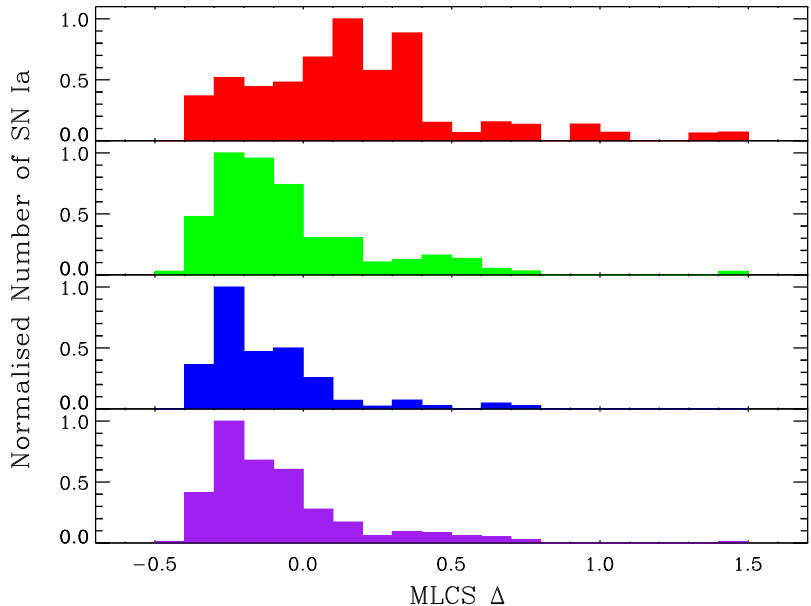}
\caption{The distribution of $\Delta$ for SNe found in passive galaxies (\emph{top panel}) is compared to those found in star-forming galaxies. The distributions for moderately star-forming and highly star-forming galaxies are plotted in the second and third panels respectively. (\emph{bottom panel:}) Star-forming galaxies are plotted as a cumulative histogram, with the distributions for moderately star-forming and highly star-forming galaxies combined.
\label{fig:snprop_delta}}
\end{figure*}

Figure~\ref{fig:snprop_delta} shows the distribution of the MLCS2k2 $\Delta$ parameter for the SNe Ia found in passive and star-forming host galaxies (shown both separately and as a combined dataset), after correcting for efficiency as described in \S\ref{sec:efficiency} and \S\ref{subsec:fieldincompleteness}. For the passive galaxies, we find a mean value of $\Delta = 0.20$, with variance $0.14$, compared to the star-forming galaxies, which have lower mean value  $\Delta = -0.08$ and a smaller variance of $0.06$. 

A Kolmogorv-Smirnov test (KS test; \citealt{chakravartistat}) and an Anderson-Darling test (AD test; \citealt{Stephens1974}) are used to test the hypothesis that the two histograms shown in Figure~\ref{fig:snprop_delta} are drawn from the same parent distribution.   We find probabilities of $2.67 \times {10}^{-13}$ for the KS test and $3.21 \times {10}^{-13}$ for the AD-test and conclude that the histograms arise from two different populations.  Our result confirms previous findings \citep{2006ApJ...648..868S,2010arXiv1005.4687L} that SNe in star-forming galaxies are brighter than their passive counter-parts and that SNe in passive galaxies exhibit a broader range of $\Delta$ values when compared to their star-forming counterparts. 

%%%%%%%%%%%%%%%%%%%%%%%%%%  Figure 6  %%%%%%%%%%%%%%%%%%%%%%%%%%%%%%
%%%%%%%%%%%%%%%%%%%%%%%%%%%%%%%%%%%%%%%%%%%%%%%%%%%%%%%%%%%%%%

To investigate further, we split the star-forming galaxy sample into moderately and highly star-forming datasets (as described in \S\ref{sec:hostgalprop}). When we compare the distribution of $\Delta$ in passive galaxies to that of moderately and highly star-forming galaxies, respectively, we find KS test probabilities of $4.8 \times {10}^{-9}$ for moderately star-forming, and $2.9 \times {10}^{-13}$ for highly star-forming galaxies, with comparable values for the AD test. This shows that SNe Ia in star-forming galaxies differ from their passive counterparts, even for moderate levels star-formation. We find a KS test probability of $0.04$ (with AD test value $0.004$) that the $\Delta$ distributions in moderately and highly star-forming galaxies arise from the same parent distribution.  

These results are not surprising in the context of a two component model since we have shown that most of the rate even in moderately star-forming galaxies can be attributed to recent star-formation. However, if there were more than two components, or some evolution depending on the star-forming rate, we might have observed a significant difference between the high star-forming and moderately star-forming distributions.

\subsection{MLCS2k2 $A_V$ Parameter as a Function of Host Galaxy Type}
\label{sec:snprop_av}

Determining the color of SNe Ia is important for cosmological parameter estimation. Recently, \citet{2010MNRAS.406..782S,2010ApJ...715..743K} and \citet{2010arXiv1005.4687L} found evidence that SNe in different environments may follow different color laws. Here we consider how the distribution of color, expressed by the MLCS2k2 $A_V$ parameter, varies as a function of host galaxy sSFR.  To examine this relationship we apply a flat prior in MLCS2k2, allowing $A_V$ to take all values (both positive and negative), so that we are not sensitive to assumptions about the distribution of $A_V$ values. The use of a flat prior changes the number of SNe Ia that pass our selection criteria from 342 to 338. The effect of our choice of prior is discussed further in \S\ref{sec:figure11_cuts}.
 
\begin{figure*}[ht]
\epsscale{1.5}
\plotone{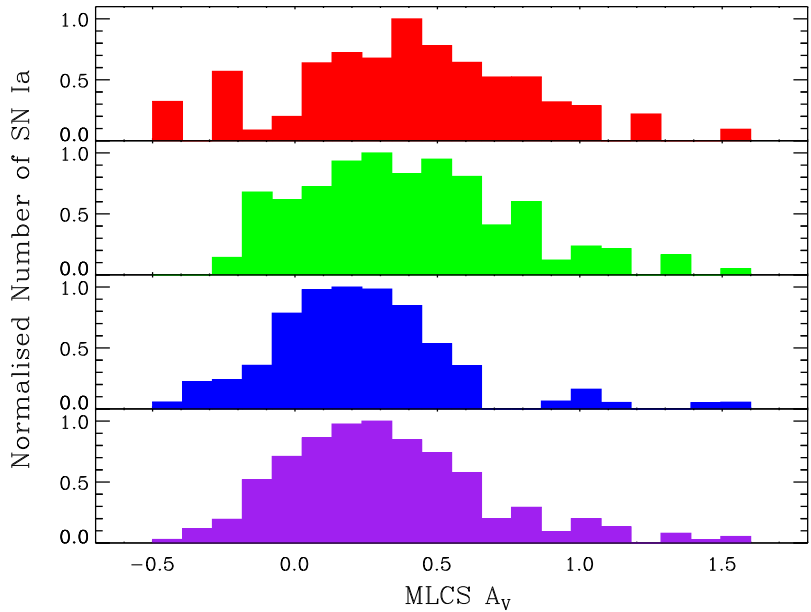}
\caption{The distribution of $A_V$ for SNe found in passive galaxies (\emph{top panel}) is compared to those found in star-forming galaxies. The distributions for moderately star-forming and highly star-forming galaxies are plotted in the second and third panels respectively. The bottom panel is the sum of the two middle panels, showing the combined distribution for star-forming galaxies. A flat prior is used in the light-curve fitting. 
\label{fig:snprop_av}}
\end{figure*}

Figure~\ref{fig:snprop_av} shows the distribution of $A_V$ for SNe in passive hosts and star-forming galaxies. The distributions for moderately star-forming and highly star-forming galaxies are plotted separately, along with the case where the two star-forming datasets have been combined. We use the efficiency correction described in \S\ref{sec:efficiency} and the $1/V_\textrm{max}$ correction determined in \S\ref{subsec:fieldincompleteness} to weight each galaxy. Passive galaxies have a mean $A_V$ of $0.40 $mag and variance $0.27$, compared to a mean of $0.33 $mag and variance $0.15$ for star-forming galaxies. For the individual star-forming galaxy populations, we find means of $0.43$ and $0.22 $mag and variances of $0.16$ and $0.12$ for moderately and highly star-forming galaxies, respectively. 

As in \S\ref{sec:snprop_delta} we use both KS and AD tests to indicate whether the distributions are drawn from the same parent distributions. We find probabilities of $0.163$ and $0.049$ respectively, suggesting no evidence that the distributions may be drawn from different parent distributions. These results are consistent with \citet{2010arXiv1005.4687L}, who used the SDSS spectroscopically confirmed SNe fitted with the SALT2 light-curve fitter, and saw no significant difference in the distribution of the SALT2 color parameter, $c$, for SNe Ia's in passive and star-forming galaxies.

%%%%%%%%%%%%%%%%%%%%%%%%%%  Figure 7  %%%%%%%%%%%%%%%%%%%%%%%%%%%%%%
%%%%%%%%%%%%%%%%%%%%%%%%%%%%%%%%%%%%%%%%%%%%%%%%%%%%%%%%%%%%%

As before, we split the star-forming dataset in to moderately and highly star-forming galaxies. We fit KS test probabilities of $0.42$ when passive and moderately star-forming datasets are considered, $8.0 \times {10}^{-4}$ for passive and highly star-forming, and $2.3 \times {10}^{-4}$ between the two star-forming datasets, and comparable AD test statistics. We conclude that whilst there is no evidence of a difference in the $A_V$ distributions between the passive and moderately star-forming datasets, the highly star-forming sample has a different distribution in $A_V$, as shown in Figure~\ref{fig:snprop_av}, with SNe Ia in highly star-forming galaxies on average exhibiting smaller values of $A_V$.  

This analysis assumes that the observed values of $A_V$ are good approximations to the true underlying values.  However, while the majority of SNe in our sample have well measured light-curves, resulting in accurate measurements of $A_V$, many of the SNe Ia in our sample have low S/N measurements. In such cases, since the underlying distribution of $A_V$ for SNe Ia is observed to be exponentially declining, the measured value of $A_V$, for an individual SNe, is more likely to be scattered towards a higher value of $A_V$ than a lower value. This would result in a higher proportion of SNe with high $A_V$ measurements compared to a distribution of SNe Ia with high S/N light-curves. 

To rigorously account for $S/N$ variations in the observations, the underlying $\Delta$ and $A_V$ distributions are determined using the method described in \citet{1995NIMPA.362..487D} and Appendix D of \citet{2009ApJS..185...32K}. To quantify the uncertainty in the underlying distributions, 60 data-sized simulations were analyzed in the same way as the data. The spread in the mean and RMS of the extracted distributions are taken to be the uncertainties in these quantities. To avoid pathologies from poorly measured photometric redshifts, only SNe Ia with a spectroscopic redshift (either from the SNe or host galaxy) are used. The resulting incompleteness was modelled in simulations and found to have a negligible impact on the results. Based on the spread in the distribution moments (both mean and RMS), we estimate that the distribution of $A_V$ in highly star-forming galaxies differs from that of passive and moderately star-forming galaxies at $3.3\sigma$ and $3.5\sigma$, respectively. 

For our sample, SNe Ia have similar S/N values at maximum brightness, for all host galaxy types. SNe Ia in passive galaxies have a mean $S/N$ of 36 (with RMS of 26) compared to means of $33$ and $42$ and variances of $17$ and $30$ for SNe Ia in moderately star-forming and highly star-forming galaxies, respectively. 

Finally, we use the SALT2 light-curve fitter \citep{2007A&A...466...11G,2010A&A...523A...7G} to determine if our results are dependent on light-curve fitting technique. SALT2 uses a color term, $c$, as a measure of the color of an individual SN Ia, but does not explicitly attribute it to dust extinction. We recover the underlying distribution of $c$ for SNe in our sample, following the technique of \citet{1995NIMPA.362..487D}, and find that the both the mean and RMS of the color distribution for SNe Ia in highly star-forming galaxies is different from those in passive and moderately star-forming galaxies at $2.7\sigma$ and $3.8\sigma$, respectively. 

A physical understanding of this difference is unclear. \citet{2010ApJ...715..986S} suggest that there may be an increased amount of dust observed in star-burst galaxies, whilst  \citet{2005ApJ...619L..39S} argue that the dust content is smaller, and covers a smaller range, in low mass, highly star-forming galaxies. 

The similarity between passive and moderately star-forming galaxies suggests that much of the spread in color could arise from variations in the explosion process or effects of the local SN environment \citep{2011MNRAS.413.3075M}.  If host galaxy dust were responsible for the difference we would expect to see less extinction in passive galaxies since they have lower dust levels  \citep{1998astro.ph..6083C}. 

\subsection{The Effect of our Selection Criteria}
\label{sec:figure11_cuts}

In \S\ref{sec:selectioncuts} we considered how the inclusion of non-spectroscopically confirmed SNe Ia and our redshift range affected the results of \S\ref{sec:massrate}, \S\ref{sec:sfrrate} and \S\ref{sec:ssfr}. Here, we carry out a similar analysis on the results of \S\ref{sec:snprop_delta} and \S\ref{sec:snprop_av}. We also consider how the $A_V$ prior used in the MLCS2k2 light-curve fits affect our conclusions. 

Tables~\ref{tab:snesystematics_delta} and~\ref{tab:snesystematics_av} show the KS test probabilities described in \S\ref{sec:snprop_delta} and \S\ref{sec:snprop_av} for various selection criteria. We consider the ``standard" $A_V$ prior discussed in \S\ref{sec:spectroissues}, a flat $A_V$ prior, as used in \S\ref{sec:snprop_av} and a prior where $A_V$ is forced to positive. We also consider the effect of varying our redshift range, and considering only spectroscopically confirmed SNe Ia. 

From Table~\ref{tab:snesystematics_delta} we see that in all cases considered, there is a very low probability that the distribution of $\Delta$ from SNe in passive galaxies matches that seen in star-forming galaxies.  Similarly, the evidence that highly and moderately star-forming galaxies are different is consistently weak.  Table~\ref{tab:snesystematics_av}, shows the KS test probabilities for the $A_V$ distributions. As in \S\ref{sec:snprop_av}, there is no evidence that the distribution of $A_V$ in passive galaxies differs from that seen in star-forming galaxies.  There is evidence that the distribution of $A_V$ in highly star-forming galaxies does not match that of passive and moderately star-forming galaxies. 

From Figure~\ref{fig:snprop_delta}, we observe that fainter, higher $\Delta$, SNe are preferentially found in passive galaxies. However, the survey efficiency considered for this analysis, is a function of redshift, and does not distinguish between SNe Ia of differing intrinsic brightness. To account for this, we consider how modelling the survey efficiency as a function of both $A_V$ and $\Delta$ affects our conclusions, and find that our results are unaffected by this additional correction.

In \S\ref{sec:snprop_av}, we used the SALT2 light-curve fitter to show that our results concerning the distribution of $A_V$ (or $c$) for SNe Ia as a function of host galaxy type are independent of light-curve fitting technique considered. SALT2 parameterizes the relationship between the peak luminosity of a SN Ia and the shape of its light-curve through the $x_1$ parameter. We recover the underlying distribution of $x_1$ for SNe in our sample, and find that both the mean and RMS of the $x_1$ distribution for SNe Ia in passive galaxies differs from those found in moderately and highly star-forming galaxies at $4.6\sigma$ and $9.6\sigma$, respectively, confirming the results of \S\ref{sec:snprop_delta}.  

We further considered the possibility that our results may depend on survey conditions or may vary as a function of position on the sky.  We split the host galaxy sample by both year and position but saw no significant effects. 

\begin{deluxetable}{cccc}
\tabletypesize{\scriptsize}
\tablecaption{Effect of our selection criteria on the results described in \S\ref{sec:sneprop} on the distribution of the light-curve decline rate parameter, $\Delta$ \label{tab:snesystematics_delta}}
\tablewidth{0pt}
\tablehead{ \colhead{Selection} & \colhead{No. Hosts} & \multicolumn{2}{c}{KS-test for $\Delta$ distribution}
}
\startdata
 &  & Passive / Star-forming & High / Mod. \\
 \hline
Std $A_V$ Prior & 342 & $2.67 \times {10}^{-13}$ & $0.037$ \\
Flat $A_V$ Prior & 338 & $6.37 \times {10}^{-12}$ & $0.040$ \\
Positive $A_V$ Prior & 364 & $3.66 \times {10}^{-14}$ & $0.037$ \\
Confirmed & 197 & $2.31 \times {10}^{-12}$ & $0.020$ \\
Phot-ID &145 & $2.17 \times {10}^{-3}$ & $0.137$ \\
$z < 0.20$ & 196 & $8.79 \times {10}^{-8}$ & $0.070$\\
$z < 0.16$ & 103 & $3.00 \times {10}^{-5}$ & $0.295$ \\
\enddata
\end{deluxetable}

\begin{deluxetable}{cccccc}
\tabletypesize{\scriptsize}
\tablecaption{Effect of our selection criteria on the results described in \S\ref{sec:sneprop} on the distribution of $A_V$ \label{tab:snesystematics_av}}
\tablewidth{0pt}
\tablehead{ \colhead{Selection} & \colhead{No. Hosts} &\multicolumn{4}{c}{KS-test for $A_V$ distribution}
}
\startdata
 &  & Passive / Star-forming & Passive / Mod. & Passive / Highs & Mod. / Highs \\
 \hline
Std $A_V$ Prior & 342 & $0.295$ & $0.889$ & $6.11 \times {10}^{-3}$ & $1.10 \times {10}^{-4}$\\
Flat $A_V$ Prior & 338 & $0.163$ & $0.416$ & $7.98 \times{10}^{-4}$ & $2.31 \times {10}^{-4}$\\
Positive $A_V$ Prior & 364 & $0.036$ & $0.923$ & $3.07 \times {10}^{-4}$ & $1.13 \times {10}^{-4}$ \\
Confirmed & 197 & $0.320$ & $0.843$ & $0.022$ & $3.13 \times {10}^{-3}$ \\
Phot-ID &145 & $0.573$ & $0.390$ & $0.271$ & $0.018$ \\
$z < 0.20$ & 196 & $0.542$ & $0.606$ & $0.034$ & $1.00 \times {10}^{-4}$ \\
$z < 0.16$ & 103 & $0.612$ & $0.032$ & $0.240$ & $8.04 \times {10}^{-5}$ \\
\enddata
\end{deluxetable}

%%%%%%%%%%%%%%%%%%%%%%%%%%%%%%%%%%%%%%%%%%%%%%%%%%%%%%%%%%%%%%%%%%%%%%%%%%%%%%%%%%

\section{Conclusions}
\label{sec:conclusions}

We have studied how the SN Ia rate and light-curve properties depend on the host galaxy stellar mass and star-formation rate. By augmenting the SDSS spectroscopically confirmed SNe Ia with SNe identified by only their light-curves, we have constructed a large, homogeneous and well-understood sample of 342 SNe Ia in the redshift range $0.05 < z < 0.25$.  Our sample has low contamination and is unbiased with respect to spectroscopic selection effects and survey conditions. The efficiency of the SDSS-II SN Survey is well measured in this redshift range, allowing us to study the overall SN Ia rate as a function of these host galaxy properties. We summarize below the main conclusions of this work:
\begin{itemize} 
\item We find that the SN Ia rate in passive galaxies is not linearly proportional to the stellar mass, but instead favoring $\snrIa \propto {M}^{0.67}$, as illustrated in Figure~\ref{fig:massrate}. This result differs from that of \citet{2006ApJ...648..868S}, at higher redshift, who favor a linear relationship, but is in good agreement with the conclusions of \citet{2010arXiv1006.4613L}, who favor $\snrIa \propto {M}^{0.487 \pm 0.316}$ for elliptical galaxies in the local Universe.  
\item For star-forming galaxies we find that the SN Ia rate as a function stellar mass differs from that of passive galaxies, instead favoring $\snrIa \propto {M}^{0.94}$. This result differs from that of \citet{2010arXiv1006.4613L}, who found that the SN Ia rate as a function of stellar mass is independent of host galaxy morphology and color. 
\item We show that the SN Ia rate per unit stellar mass is a strong function of specific star-formation rate (sSFR), with SNe Ia being preferentially found in highly star-forming or starburst galaxies, compared to their passive counterparts (Figure~\ref{fig:ssfr}). This relationship is consistent with those found by  \citet{2005A&A...433..807M} and \citet{2006ApJ...648..868S}, locally and at high redshift, respectively, implying that this relationship does not evolve with redshift. 
\item We demonstrate that the excess SN Ia rate in star-forming galaxies is well fit by a linear relationship proportional to the recent star-formation rate, as shown in Figure~\ref{fig:sfrrate}. The component related to recent star-formation is the dominant contributor to the SN Ia rate in these galaxies. 
\item We find that a bivariate fitting technique confirms that SNe Ia in this sample satisfy a SN Ia rate of the form $\snrIa = {1.1 \pm 0.2 \times {10}^{-10} \mgal^{0.7 \pm 0.01}} + 1.0 \pm 0.1 \times {10}^{-3} {\dot{M}^{1.0 \pm 0.1}}$  (statistical errors only). 
This parameterization is a generalization of the A+B model and provides a better fit to the SDSS-II SN data than assuming a SN Ia rate linearly dependent on stellar mass and star-formation rate.
\item We have tested the effect of our selection criteria on these results, and find that the exclusion of  photometrically classified SNe Ia's, and variations in redshift range, do not significantly alter our results. 
\item We confirm the striking difference in light-curve shape between passive and star-forming galaxies.  Specifically, brighter, slowly declining SNe (with smaller $\Delta$ values for MLCS2k2) are seen preferentially in star-forming galaxies while faint, quickly declining SNe (with high $\Delta$ values) are preferentially found in passive galaxies as shown in Figure~\ref{fig:snprop_delta}. 
\item We see no difference in the distribution of the extinction parameter, $A_V$, between passive and star-forming galaxies as illustrated in Figure~\ref{fig:snprop_av}. We find no evidence that the distribution of $A_V$ in passive galaxies differs from that of moderately star-forming galaxies, but find evidence that the distribution is different for highly star-forming galaxies, which favor lower mean values of $A_V$. We use the method described in Appendix D of \citet{2009ApJS..185...32K} to determine the underlying distribution of $A_V$ for various galaxy types, and show that the distribution of $A_V$ in highly star-forming galaxies differs at the $3\sigma$ level from that of passive and moderately star-forming galaxies.  We find that the choice of $A_V$ prior used in the light-curve fitting does not affect our conclusions.  We find the same results using the SALT2 model to extract the distribution of color, $c$, thus providing evidence that the difference in the distribution of $A_V$ (or $c$) is a model-independent feature.
\item We perform a rigorous test of the \PEGASE SEDs and the Z-PEG fitting technique and find a systematic offset in the photometric redshift estimates produced for our comparison field sample. If a simple correction is applied to both the redshifts and stellar masses of our field sample, we find that our conclusions are unchanged. 
\end{itemize} 
The process used to determine our sample of host galaxies and their derived properties allows us to directly compare our conclusions with those of \citet{2006ApJ...648..868S}, who studied the SN Ia rate at higher redshifts. \citet{2006ApJ...648..868S} found the same trends with star-formation rate, but with a different relationship parameterizing the stellar mass into the SN Ia rate. It is unlikely that these differences are due to an evolution of the galaxy population but may reflect that SNe Ia are primarily triggered by recent bursts of star-formation in a galaxy, causing uncertainties in the contribution due to the stellar mass.

\acknowledgments

\section*{Acknowledgements}

The authors thank Mark Sullivan, Damien Le Borgne, Claudia Maraston and Janine Pforr for helpful discussions. MS is funded by an SKA fellowship, while RCN and HL are supported by STFC. MS thanks Prina Patel and Russell Johnston for insightful comments. Support for this research at Rutgers University was provided in part by NSF CAREER award AST-0847157 to SWJ.

Funding for the SDSS and SDSS-II has been provided by the Alfred P. Sloan Foundation, the Participating Institutions, the National Science Foundation, the U.S. Department of Energy, the National Aeronautics and Space Administration, the Japanese Monbukagakusho, the Max Planck Society, and the Higher Education Funding Council for England. The SDSS Web Site is http://www.sdss.org/.

The SDSS is managed by the Astrophysical Research Consortium for the Participating Institutions. The Participating Institutions are the American Museum of Natural History, Astrophysical Institute Potsdam, University of Basel, University of Cambridge, Case Western Reserve University, University of Chicago, Drexel University, Fermilab, the Institute for Advanced Study, the Japan Participation Group, Johns Hopkins University, the Joint Institute for Nuclear Astrophysics, the Kavli Institute for Particle Astrophysics and Cosmology, the Korean Scientist Group, the Chinese Academy of Sciences (LAMOST), Los Alamos National Laboratory, the Max-Planck-Institute for Astronomy (MPIA), the Max-Planck-Institute for Astrophysics (MPA), New Mexico State University, Ohio State University, University of Pittsburgh, University of Portsmouth, Princeton University, the United States Naval Observatory, and the University of Washington.

This work is based in part on observations made at the following telescopes. The Hobby-Eberly Telescope (HET) is a joint project of the University of Texas at Austin, the Pennsylvania State University, Stanford University, Ludwig-Maximillians-Universit\"{a}t M\"{u}nchen, and Georg-August-Universit\"{a}t G\"{o}ttingen. The HET is named in honor of its principal benefactors, William P. Hobby and Robert E. Eberly. The Marcario Low-Resolution Spectrograph is named for Mike Marcario of High Lonesome Optics, who fabricated several optical elements for the instrument but died before its completion; it is a joint project of the Hobby-Eberly Telescope partnership and the Instituto de Astronom\'{i}a de la Universidad Nacional Aut\'{o}noma de M\'{e}xico. The Apache Point Observatory 3.5 m telescope is owned and operated by the Astrophysical Research Consortium. We thank the observatory director, Suzanne Hawley, and site manager, Bruce Gillespie, for their support of this project. The Subaru Telescope is operated by the National Astronomical Observatory of Japan. The William Herschel Telescope is operated by the Isaac Newton Group, on the island of La Palma in the Spanish Observatorio del Roque de los Muchachos of the Instituto de Astrofisica de Canarias. Based on observations made with the Nordic Optical Telescope, operated on the island of La Palma jointly by Denmark, Finland, Iceland, Norway, and Sweden, in the Spanish Observatorio del Roque de los Muchachos of the Instituto de Astrofisica de Canarias. Kitt Peak National Observatory, National Optical Astronomy Observatory, is operated by the Association of Universities for Research in Astronomy, Inc. (AURA) under cooperative agreement with the National Science Foundation. The W.M. Keck Observatory is operated as a scientific partnership among the California Institute of Technology, the University of California, and the National Aeronautics and Space Administration. The Observatory was made possible by the generous financial support of the W. M. Keck Foundation. Based partially on observations made with the Italian Telescopio Nazionale Galileo (TNG) operated on the island of La Palma by the Fundaci\'{o}n Galileo Galilei of the INAF (Istituto Nazionale di Astrofisica) at the Spanish Observatorio del Roque de los Muchachos of the Instituto de Astrof\'{i}sica de Canarias.

\appendix

\section{Color-Magnitude Diagram}
\label{app:color-mag}

\begin{figure}[ht]
\epsscale{.80}
\plotone{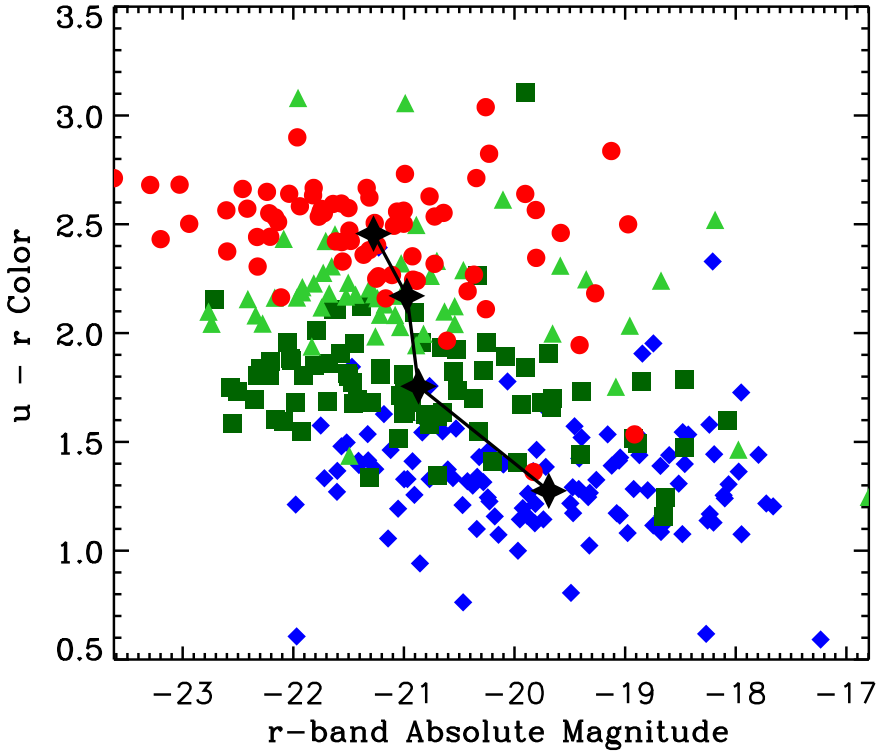}
\caption{Color (\textit{u-r}) versus absolute magnitude for the host galaxy sample used in this analysis. Galaxies classified as passive are plotted as red circles, with highly star-forming galaxies shown as blue diamonds. Moderately star-forming galaxies are plotted in green, with light-green triangles indicating the ``ridge-line" of galaxies discussed in \S\ref{sec:hostgalprop} and shown in Figure~\ref{fig:massvsfr}. The mean of each individual distribution is shown as a black star, to indicate the relationship between color and brightness for the various samples. 
 \label{fig:app:colormag}}
\end{figure}

In \S\ref{sec:hostgalprop} we considered how our host galaxies are distributed as a function of stellar mass and star-formation rate, and noted the presence of a subset of galaxies with low-levels of specific star-formation rate. These galaxies do not exhibit high $\chi^2$ values. Figure~\ref{fig:app:colormag} shows the color-magnitude diagram for the host galaxies used in this analysis. The sample is split into passive, moderately and highly star-forming as described in the caption for Figure~\ref{fig:massvsfr}. The subset of 55 galaxies with low levels of sSFR ($-11.0 < \textrm{sSFR} < 10.5$) are plotted as light-green triangles. We see that the \PEGASE SED primarily determine the level of star-formation activity in a galaxy based on its color, with passive galaxies being the ``reddest" and brightest galaxies through to the ``bluest" galaxies being classified as highly star-forming. From Figure~\ref{fig:app:colormag}, the population of objects with low sSFR but classified as moderately star-forming are observed to lie between the passive and the other moderately star-forming galaxies in color-magnitude space, making their classification understandable, if ambiguous. Figure~\ref{fig:app:colormag} uses absolute magnitudes and colors, but the same conclusions can be drawn when apparent magnitudes are considered. We note that, 43 of the 55 ($78\%$) ``ridge line" galaxies are best described by the lenticular (S0) scenario, with the remaining 12 being best-fit by the elliptical galaxy template, possibly highlighting the uncertainty in the nature of lenticular galaxies. In comparison, only 41 of the 79 ($52\%$) remaining moderately star-forming galaxies are best-fit by an S0 template. 

To further investigate the nature of the ``ridge-line" galaxies, we study the how distribution of $\Delta$ for SNe in these galaxies compare to those in moderately star-forming and passive galaxies, since, as described in \S\ref{sec:snprop_av}, there is a significant difference in the $\Delta$ distributions for SN Ia occurring in these galaxies. A K-S test between the distribution of $\Delta$ in passive galaxies to that of ``ridge-line galaxies" yields a probability of $1.9\times {10}^{-5}$ that they are drawn from the same parent distribution. This compares to a probability of $0.28$ between the distribution of $\Delta$ in ``ridge-line" galaxies and other moderately star-forming galaxies. These results, further strengthen the conclusion that the ``ridge-line" galaxies are not misclassified passive galaxies. 

%%%%%%%%%%%%%%%%%%%%%%%%%%  Figure 8  %%%%%%%%%%%%%%%%%%%%%%%%%%%%%%
%%%%%%%%%%%%%%%%%%%%%%%%%%%%%%%%%%%%%%%%%%%%%%%%%%%%%%%%%%%%%%

We thus determine that this population of objects is well defined according to color-magnitude space, and that the \PEGASE SEDs use this information to determine the level of star-formation activity in each host galaxy. 

\section{The \PEGASE Photometric Redshifts}
\label{app:pegase_photoz}

One of the key systematic uncertainties in this analysis concerns the accuracy of the derived properties of the comparison field sample used, and in particular the photometric redshift estimates produced by the \PEGASE SEDs. These redshift estimates will affect not only the number of field galaxies, but also their associated stellar masses. The photometric redshift estimates have been tested at high redshift by \citet{2006ApJ...648..868S}, but have not been extensively used in the local Universe. 

\begin{figure}[ht]
\epsscale{1.0}
\plotone{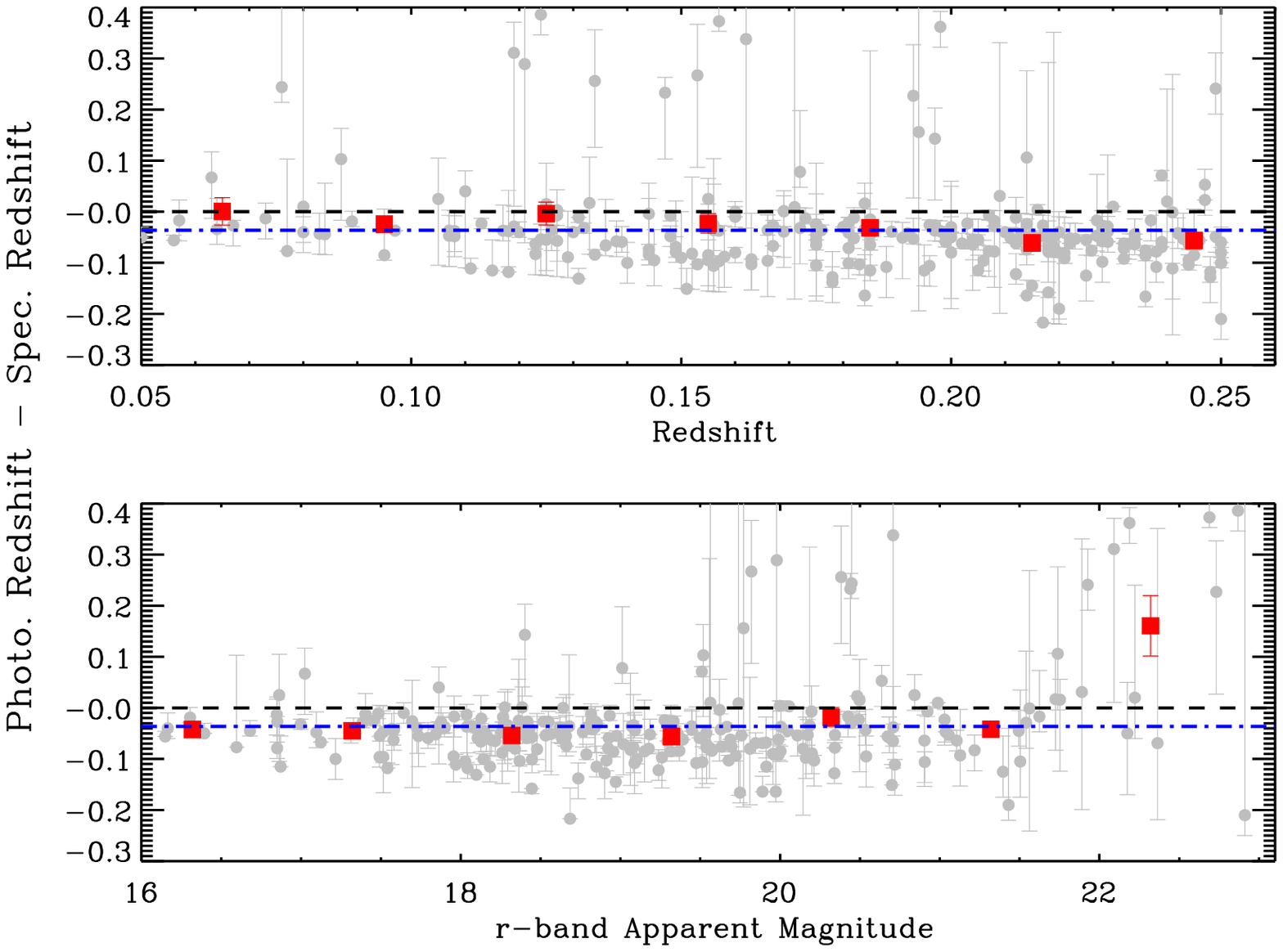}
\caption{\emph{Top:} The difference between the photometric redshift estimates derived from the \PEGASE SEDs and the spectroscopic redshift as a function of redshift. \emph{Bottom:} Same as above, as a function of host galaxy apparent magnitude. Individual galaxies are plotted in grey, with red points indicating the values determined when the sample has been binned. The black, dashed line indicates no difference, while the blue (dashed-dotted) line indicates the mean difference.  \label{fig:app:photoz_redshift}}
\end{figure}

To test the accuracy of the photometric redshifts, we use the host galaxy sample described in \S\ref{sec:galaxydata}, and whose properties are listed in Table~\ref{tab:listofobjects}. This sample covers the magnitude range of the comparison field sample and is large enough to statistically determine if the photometric estimates are accurate. Figure~\ref{fig:app:photoz_redshift} shows the difference between the photometric redshift estimates and the known spectroscopic redshift for this sample as a function of both redshift and apparent magnitude. We find a mean difference of $0.03$ in redshift, with the photometric redshifts being smaller than the spectroscopic redshift. From Figure~\ref{fig:app:photoz_redshift}, there is no evidence of this offset being dependent on either the redshift or apparent magnitude of the host galaxy, although the scatter does increase with apparent magnitude.

%%%%%%%%%%%%%%%%%%%%%%%%%%  Figure 9  %%%%%%%%%%%%%%%%%%%%%%%%%%%%%%
%%%%%%%%%%%%%%%%%%%%%%%%%%%%%%%%%%%%%%%%%%%%%%%%%%%%%%%%%%%%%

\begin{figure}[ht]
\epsscale{1.0}
\plotone{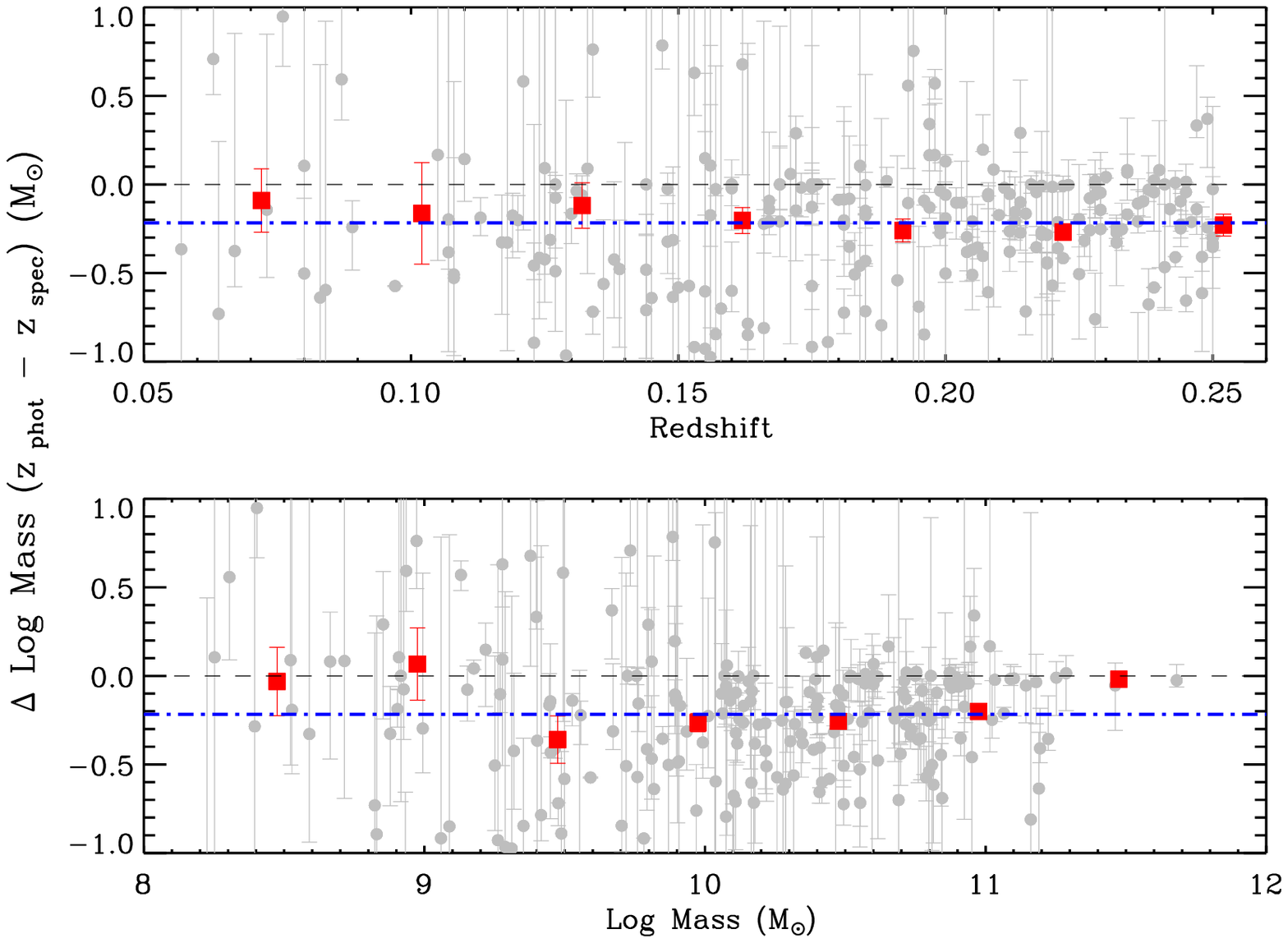}
\caption{\emph{Top:} The difference between the stellar mass derived when the redshift is allowed to float in the Z-PEG code compared to that when the redshift is held fixed, resulting in the offset described in Figure~\ref{fig:app:photoz_redshift}, as a function of redshift. \emph{Bottom:} Same as above, as a function of stellar mass, as derived when the redshift is held fixed. Individual galaxies are plotted in grey, with red points indicating the values determined when the sample has been binned. The black, dashed line indicates no difference, while the blue (dashed-dotted) line indicates the mean difference. \label{fig:app:photoz_mass}}
\end{figure}

This observed offset in the photometric redshift will also lead to an incorrect value for the galaxy's stellar mass. To quantify this, we consider the derived stellar mass when the redshift is held fixed, compared to that when it is allowed to float in Figure~\ref{fig:app:photoz_mass}, resulting in the offset described above. An offset of $\log M = 0.22$ is seen, with the stellar masses derived when the redshift is allowed to float being smaller than the value determined when the redshift is known.

%%%%%%%%%%%%%%%%%%%%%%%%%%  Figure 10  %%%%%%%%%%%%%%%%%%%%%%%%%%%%%%
%%%%%%%%%%%%%%%%%%%%%%%%%%%%%%%%%%%%%%%%%%%%%%%%%%%%%%%%%%%%%%

\section{The Effect of the Observed Offset on the Conclusions of this Work}
\label{app:effect} 

The offset between the photometric redshifts produced by the \PEGASE SEDs and the spectroscopic redshifts for the host galaxy sample implies that the distribution of galaxies in the comparison field sample used in this analysis do not accurately reflect the distribution of galaxies in our redshift range. This may affect the results of \S\ref{sec:hostrate}. Here we attempt to quantify this systematic uncertainty. 

In Appendix~\ref{app:color-mag}, we observed that there is a strong dependence between the color of the host galaxy and the best-fitting \PEGASE template determined by the Z-PEG code. This is true for colors determined both by using absolute and apparent magnitudes. The reddest galaxies (in $u-r$) are well-fit by a passive template, through to the bluest galaxies, which are considered to be highly star-forming. Thus, it appears that, we can approximately describe the level of star-formation inferred by the \PEGASE templates as purely a function of observed quantities, and not affected by the offset described in Appendix \ref{app:pegase_photoz}. We hence assume, for this analysis, that the offset in the photometric redshifts determined from the \PEGASE SEDs in our redshift range purely affect the inferred stellar masses and not the star-formation rates. We note that this approximation will only be valid for galaxies spanning a narrow redshift range, as large relative k-correction terms can lead to a color dependence, affecting the relationship determined in Figure~\ref{fig:app:colormag}.

In Appendix \ref{app:pegase_photoz} we showed that there is no evidence that the difference in redshift ($|z_{\textrm{photo}} - z_{\textrm{spec}}|$) and difference in stellar mass ($|\log M(z_{\textrm{photo}}) - \log M(z_{\textrm{spec}})|$) are dependent on either redshift or apparent magnitude. We thus assume that the photometric redshifts derived from the \PEGASE SEDs and associated stellar masses can be offset by the values determined in Appendix \ref{app:pegase_photoz}. Table~\ref{tab:app:effect} shows the effect that correcting the redshifts and stellar masses of the comparison field sample has on several of the key parameters discussed in this work, when the differences found in Appendix \ref{app:pegase_photoz} are applied.

\begin{deluxetable}{ccc}
\tablecaption{Table showing how the SN rate parameters determined in this paper are altered when the offsets in redshift and stellar mass, as determined in Appendix~\ref{app:pegase_photoz}, are applied to the comparison field sample} 
\tablewidth{0pt}
\tablehead{  \colhead{Parameter} & \colhead{Original Result} & \colhead{Mean Offset} \label{tab:app:effect}}
\startdata
No. Field Galaxies $^a$ & 733688 & 615906 \\
Total Stellar Mass of Field Galaxies $^{a,b}$ & $6.67$ & $9.50$ \\
Passive Rate $^c$ & $3.56 \pm 0.45$ & $2.49 \pm 0.31$ \\
$\nmass$  $^d$ & $0.680 \pm 0.150 $ & $0.445 \pm 0.120$ \\
$\nsfr$ $^e$& $0.940 \pm 0.078$ & $0.782 \pm 0.061$ \\
$\nsfr$ (when $\nmass=1$)  $^f$& $0.987 \pm 0.081$ & $1.070 \pm 0.075$ \\
$\nsfr$ (when $\nmass \neq1$)  $^g$& $0.955 \pm 0.074$ & $1.041 \pm 0.070$\\
\enddata
\tablenotetext{a}{After the magnitude cut ($15.5 < \textit{r} < 23.0$) and redshift cut ($0.05 < z < 0.25$)}
\tablenotetext{b}{In units of $1\times {10}^{15}$\msun}
\tablenotetext{c}{The SN rate per unit stellar mass per year in passive galaxies, as described in \S\ref{sec:ssfr}, in units of $1\times {10}^{-14}$ per unit stellar mass per year}
\tablenotetext{d}{The SN rate per galaxy per year for passive galaxies as a function of log stellar mass, as described in \S\ref{sec:massrate}}
\tablenotetext{e}{As \textit{d}, except for all star-forming galaxies combined, as described in \S\ref{sec:massrate}}
\tablenotetext{f}{The SN rate per galaxy per year for star-forming galaxies as a function of log star-formation rate, after assuming a component proportional to the stellar mass, as described in \S\ref{sec:sfrrate}}
\tablenotetext{g}{As \textit{f}, only assuming a component proportional to the values determined in \textit{d}, as described in \S\ref{sec:sfrrate}}
\end{deluxetable}

From Table~\ref{tab:app:effect}, it is clear that the number of field galaxies in the redshift range, $0.05 < z < 0.25$, is dramatically reduced when this corrections is applied, but there is also an increase in stellar mass of each galaxy, resulting in the total stellar mass of the field sample being increased from when no correction is applied. Consequently, the SN Ia rate per unit stellar mass per year in passive galaxies is decreased (at the $2.4\sigma$ level, when only statistical errors are considered) when the correction is made. This is still in good agreement with other measurements \citet{2006ApJ...648..868S,2005A&A...433..807M}. 

When considering the effect that the offset in redshift and stellar mass has on the exponents considered for the SN Ia rate, we see that the value of the slope determined in \S\ref{sec:massrate} decreases, although only by $1.6\sigma$. In both cases, a linear relationship between the SN Ia rate per galaxy per year and stellar mass in passive galaxies is strongly disfavored, confirming the results of \S\ref{sec:massrate}. 

There is a corresponding decrease in the relationship between SN Ia rate and stellar mass for star-forming galaxies (at the $2.0\sigma$ level), although as in the main result, there is a clear difference between the rate in passive galaxies when compared to that in star-forming galaxies, indicating the need for a SN Ia rate that is dependent on more than stellar mass. Finally, we consider the results of \S\ref{sec:sfrrate}. These results are statistically unaffected by the offset discovered in Appendix \ref{app:pegase_photoz}. We thus conclude that the main results of this work, namely that the SN Ia in passive galaxies is not linearly related to the stellar mass, and that the SN Ia rate in star-forming galaxies is dominated by any recent burst of star-formation, are not dependent on issues surrounding the ability of \PEGASE to accurately determine the photometric redshifts for our comparison field sample. These corrections are not applied in our analysis as a clear understanding of the cause of this offset has not been found, and thus we have only been able to estimate the magnitude of it's effect on our results. 

\section{The \PEGASE Stellar Mass and SFR Estimates}
\label{app:brinc}

In Appendix \ref{app:pegase_photoz} we compared the \PEGASE photometric redshifts to a similarly distributed sample of galaxies with spectroscopic redshifts, and determined a bias in both redshift and stellar mass. Here we consider how our derived properties from the \PEGASE SEDs compare to those determined from the spectral features of a sample of SDSS-I galaxies. \citet{2003MNRAS.341...33K} and \citet{2004MNRAS.351.1151B} used the $4000\text{\AA}$ break and the Balmer absorption line index ($H {\delta}_{A}$) to measure the stellar masses and instantaneous star-formation rates for galaxies in the SDSS-I DR4 spectroscopic catalog \citep{2006ApJS..162...38A}. 

While a comparison between the stellar mass and star-formation rates determined by the \PEGASE templates and the results of \citet{2003MNRAS.341...33K} and \citet{2004MNRAS.351.1151B} may have limitations (this sample consists of only the brightest galaxies in our host galaxy sample, the resolution of the SDSS-I spectra is not optimal, and this method relies on the same underlying physics as the \PEGASE templates), it provides a useful validation of the \PEGASE measurements. We use $\sim 330,000$ galaxies from the SDSS-I catalog, limiting ourselves to the redshift range considered in this analysis. Several spectral measurements are available; we use the dust-corrected stellar mass (median value) and total star-formation rate (median of the likelihood distribution), and determine estimates from the \PEGASE SEDs using the spectroscopic redshifts and model magnitudes. 

\begin{figure}[ht]
\epsscale{1.0}
\plotone{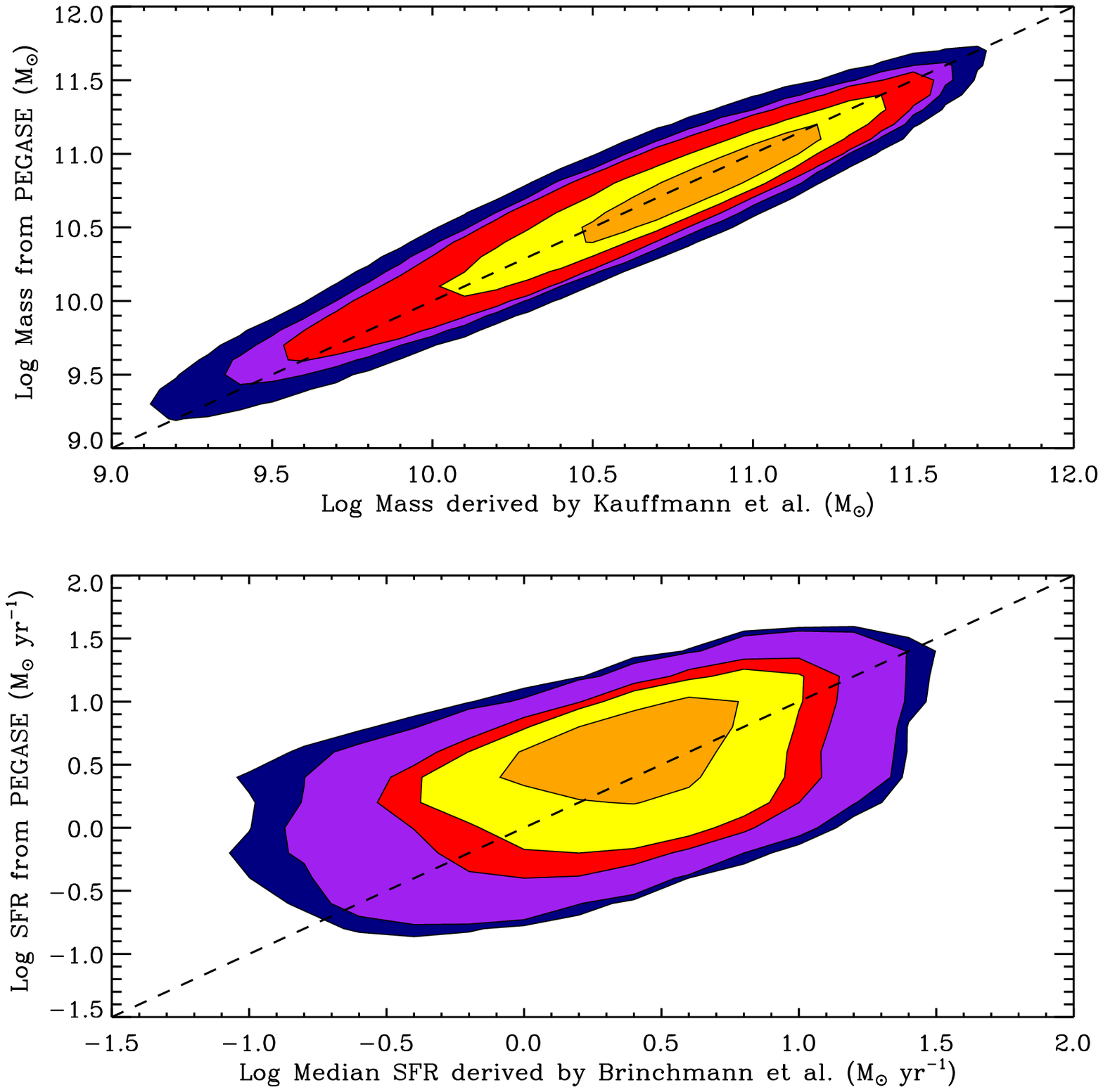}
\caption{\emph{Top:} Log stellar mass derived from the \PEGASE templates for a sample of $\sim 330,000$ galaxies from SDSS-I, compared to the estimates obtained from spectral features by \citet{2003MNRAS.341...33K}  \emph{Bottom:} Same as above, comparing \PEGASE star-formation rates to those of  \citet{2004MNRAS.351.1151B}. Contours enclose $99\%$ (dark blue), $95\%$ (purple), $90\%$ (red) $68\%$ (yellow) and $35\%$ (orange) of the data, respectively. \label{fig:app:brinc_test}}
\end{figure}

Figure~\ref{fig:app:brinc_test} shows the relationship between the spectroscopic and photometric estimates of stellar mass and star-formation rate. The total stellar mass is well recovered, with a mean offset of only $\Delta \log M = 0.001$, and variance $0.028$, well below the range significant for our analysis. 

No variation is seen with redshift or apparent magnitude. The relationship with star-formation rate shows greater scatter. We find a mean difference of $\log \textrm{SFR} = 0.115$ and variance $0.216$, with estimates from \PEGASE being larger. This is not surprising, primarily as \PEGASE estimates the mean star-formation rate averaged over the last $0.5\textrm{Gyr}$, while \citet{2004MNRAS.351.1151B} attempts to determine the instantaneous star-formation rate. 

Nevertheless, the \PEGASE and \citet{2004MNRAS.351.1151B} values agree to within $30\%$ while the star-forming rates span over two orders of magnitude. No variation is seen as a function of redshift or apparent magnitude. We note that this analysis has primarily considered passive and star-forming galaxies separately, which does not require an accurate measurement of the star-forming rate.  

%%%%%%%%%%%%%%%%%%%%%%%%%%  Figure 11  %%%%%%%%%%%%%%%%%%%%%%%%%%%%%%
%%%%%%%%%%%%%%%%%%%%%%%%%%%%%%%%%%%%%%%%%%%%%%%%%%%%%%%%%%%%%%

\section{Rest Wavelength Coverage}
\label{app:restwave} 

Throughout this analysis we have compared our observations at $z<0.25$ to those at higher redshift \citep{2006ApJ...648..868S}. While our methodology is identical to that used by \citet{2006ApJ...648..868S}, both analyses use filter sets that cover the same observed wavelength range, and thus the \PEGASE SED fits are carried out over different rest-wavelength ranges. To test how this difference may affect our conclusions, specifically the comparison to \citet{2006ApJ...648..868S}, we consider how carrying out our \PEGASE fits using a reduced number of filters affects the derived stellar mass and star-formation rates. 

The analysis of \citet{2006ApJ...648..868S} covers a bluer part of the rest-wavelength spectrum than our analysis. Thus, to produce a combination of filters that closely mimics their work, we use a reduced number of filters, removing the reddest bands from the fitting process. Specifically we investigate the cases where only the \textit{ugr} and \textit{ugri} filters are used. 

\begin{figure}[ht]
\epsscale{1.0}
\plotone{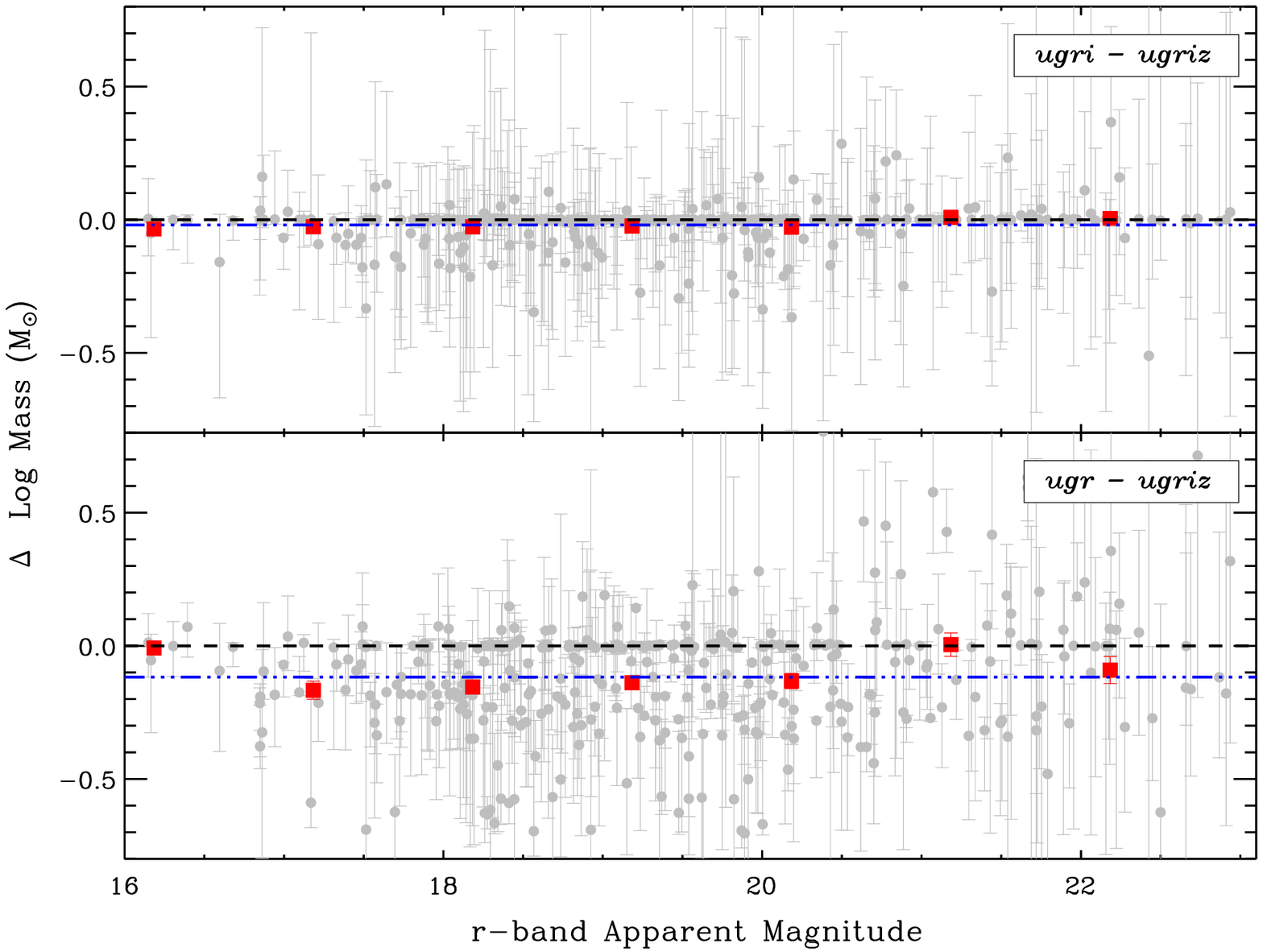}
\caption{\emph{Top:} The stellar mass derived from the \PEGASE SEDs when all five SDSS filters \textit{ugriz} are used in fit compared to when only \textit{ugri} are used as a function of apparent magnitude.  \emph{Bottom:}  Same as above, except only three filters (\textit{ugr}) are used. Individual galaxies are plotted in grey, with red squares indicating the values determined when the sample has been binned. The black, dashed line indicates no difference, while the blue (dashed-dotted) line indicates the mean difference.  \label{fig:app:bandstest_mass}}
\end{figure}

\begin{figure}[ht]
\epsscale{1.0}
\plotone{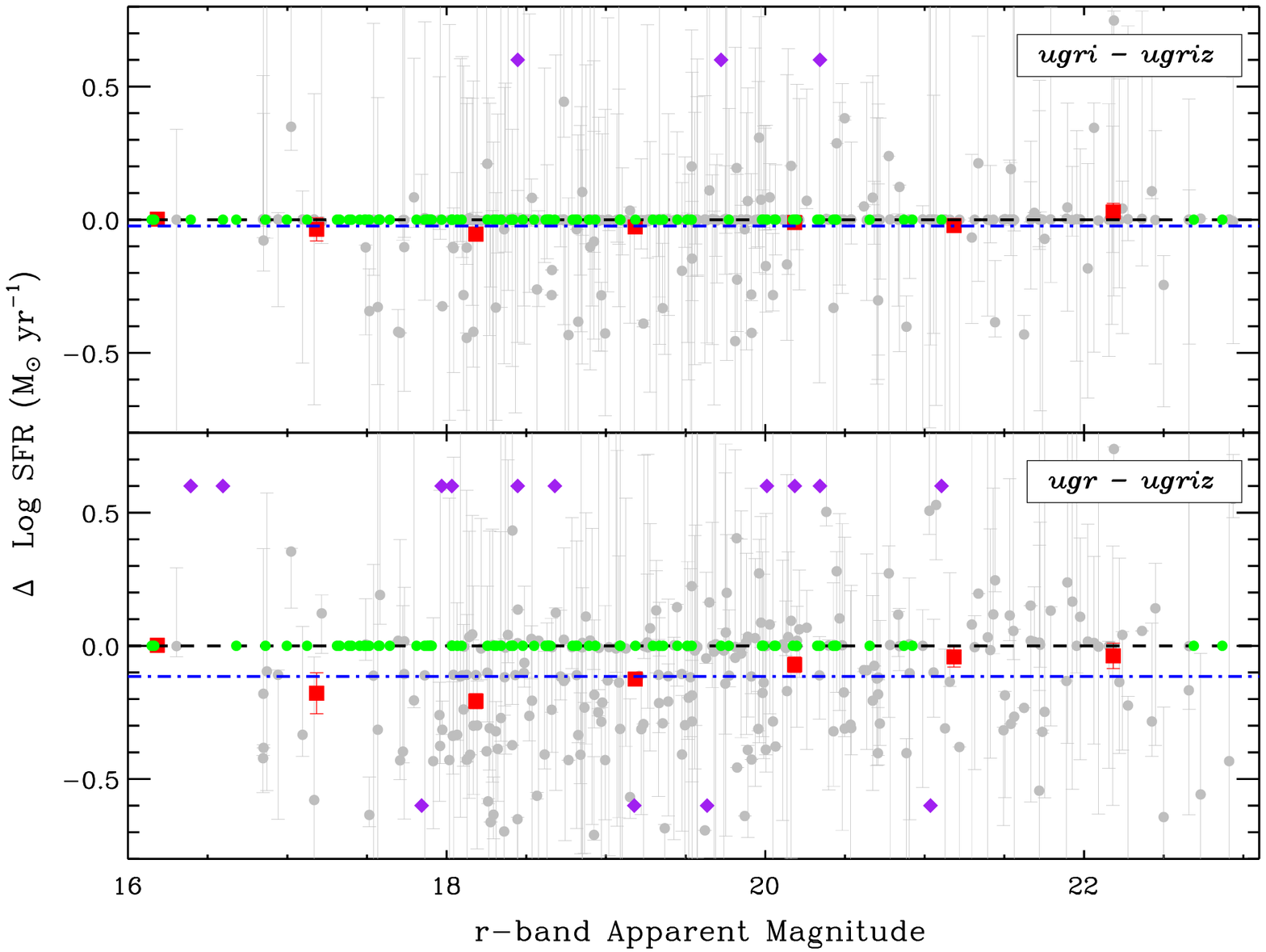}
\caption{\emph{Top:} The star-formation rate derived from the \PEGASE SEDs when all five SDSS filters \textit{ugriz} are used in fit compared to when only \textit{ugri} are used as a function of apparent magnitude.  \emph{Bottom:}  Same as above, except only three filters (\textit{ugr}) are used. Individual star-forming galaxies are plotted in grey, with red points indicating the values determined when these galaxies have been binned. Green points (plotted at $\Delta \log(\textrm{SFR}) =0$) indicate those that are determined to be passive in both cases, while purple diamonds (shown here at $\pm 0.6$) are those which are determined to be star-forming in only one scenario. The black, dashed line indicates no difference, while the blue (dashed-dotted) line indicates the mean difference.  \label{fig:app:bandstest_sfr}}
\end{figure}

Figures~\ref{fig:app:bandstest_mass} and ~\ref{fig:app:bandstest_sfr} show the difference, as a function of apparent magnitude, between the stellar masses and star-formation rates derived by the \PEGASE SEDs when various filter combinations are used. As the number of filters is reduced, and thus the number of data points used in the \PEGASE fits is reduced, the scatter between the stellar mass and star-formation rate distribution is increased. However, in the case where 4 filters are considered, no significant offset is seen, with a mean difference of $\log M = -0.02$, where galaxies are determined to be slightly less massive when only four filters are used. The star-formation rates are well recovered, with a mean difference of $\log \mathrm{SFR} = -0.024$, and scatter $\sigma = 0.141$. There is no evidence that this result is dependent on the magnitude of the galaxy. Altering the number of filters used in the \PEGASE fits allows each galaxy to be classified differently. Of the 85 galaxies that were considered passive when five filters are used, only three ($3.6\%$) are classified as star-forming, when the \textit{z}-band is omitted.

When only three filters are used to determine the derived parameters, the observed scatter increases as expected for both the stellar mass and star-formation rate distributions. A mean difference of $\log M = -0.12$ with scatter $\sigma = 0.23$ is seen. For the star-formation rate distribution, a scatter with mean difference $\log \textrm{SFR} = -0.12$ and $\sigma = 0.23$ is found. ten galaxies ($11.8\%$) that were considered to be passive when all five filters were considered are classified as star-forming when only the \textit{ugr} filters are used in the \PEGASE fits. Four galaxies ($1.5\%$) which were previously classified as star-forming are determined to be passive when only three filters are used. No trend with apparent magnitude is evident in either case. 

%%%%%%%%%%%%%%%%%%%%%%%%%%  Figure 12  %%%%%%%%%%%%%%%%%%%%%%%%%%%%%%
%%%%%%%%%%%%%%%%%%%%%%%%%%%%%%%%%%%%%%%%%%%%%%%%%%%%%%%%%%%%%%

%%%%%%%%%%%%%%%%%%%%%%%%%%  Figure 13  %%%%%%%%%%%%%%%%%%%%%%%%%%%%%%
%%%%%%%%%%%%%%%%%%%%%%%%%%%%%%%% %%%%%%%%%%%%%%%%%%%%%%%%%%%%%%

Since the extra information from the \textit{i} and \textit{z} filters does not seem to significantly cause an offset in the \PEGASE fits to higher or lower stellar masses or star-formation rates, it appears that the rest-wavelength coverage does not affect our comparisons to \citet{2006ApJ...648..868S}. 

\clearpage

\begin{deluxetable}{ccccccccc}
\tabletypesize{\scriptsize}
\tablecaption{List of SN Ia Events used in this paper and their associated host galaxy properties. \label{tab:listofobjects}}
\tablewidth{0pt}
\tablehead{
\multicolumn{2}{c}{designation} &
\multicolumn{2}{c}{host position} &
\multicolumn{1}{c}{redshift} &
\multicolumn{1}{c}{stellar mass} &
\multicolumn{1}{c}{SFR $^a$} & \multicolumn{1}{c}{sSFR $^a$} & \multicolumn{1}{c}{SN $^b$} \\
\colhead{SN ID} & \colhead{IAU} & \colhead{$\alpha(J2000)$} & \colhead{$\delta(J2000)$} & \colhead{ } & \colhead{$[\log M_{\odot}]$} & \colhead{$[\log M_{\odot}/yr]$} & \colhead{$[{yr}^{-1}]$} & \colhead{ } }
\startdata
762 & 2005eg & ${01}^{h}{02}^{m}{08.650}^{s}$ & $-{00}^{\circ}{52}^{`}{46.766}^{``}$ & 0.1915$\pm$0.0001 & $11.02^{+0.01}_{-0.18}$ & $0.27 < 0.38 < 0.71$ & -10.64 & sn \\
779 & N/A & ${01}^{h}{46}^{m}{41.703}^{s}$ & $-{01}^{\circ}{01}^{`}{14.270}^{``}$ & 0.2377$\pm$0.0005 & $10.10^{+0.01}_{-0.01}$ & $0.92 < 0.93 < 0.93$ & -9.18 & gal \\
822 & N/A & ${02}^{h}{42}^{m}{14.579}^{s}$ & $-{00}^{\circ}{51}^{`}{43.607}^{``}$ & 0.2166$\pm$0.0183 & $9.85^{+0.02}_{-0.19}$ & $-99.00 < -0.79 < -0.46$ & -10.64 & lc \\
911 & N/A & ${02}^{h}{34}^{m}{45.829}^{s}$ & $-{00}^{\circ}{06}^{`}{54.968}^{``}$ & 0.2080$\pm$0.0100 & $10.29^{+0.07}_{-0.29}$ & $0.59 < 0.80 < 0.81$ & -9.49 & gal \\
1008 & 2005il & ${01}^{h}{53}^{m}{06.704}^{s}$ & $+{01}^{\circ}{06}^{`}{49.642}^{``}$ & 0.2260$\pm$0.0100 & $10.46^{+0.02}_{-0.18}$ & $-99.00 < -0.18 < 0.15$ & -10.64 & gal \\
1032 & 2005ez & ${03}^{h}{07}^{m}{11.018}^{s}$ & $+{01}^{\circ}{07}^{`}{11.982}^{``}$ & 0.1297$\pm$0.0002 & $10.47^{+0.10}_{-0.07}$ & $-99.00 < -99.00 < -99.00$ & -99.00 & sn \\
1241 & 2005ff & ${22}^{h}{30}^{m}{41.147}^{s}$ & $-{00}^{\circ}{46}^{`}{34.472}^{``}$ & 0.0902$\pm$0.0005 & $10.53^{+0.18}_{-0.21}$ & $-99.00 < -99.00 < -99.00$ & -99.00 & sn \\
1371 & 2005fh & ${23}^{h}{17}^{m}{29.700}^{s}$ & $+{00}^{\circ}{25}^{`}{46.826}^{``}$ & 0.1191$\pm$0.0001 & $10.76^{+0.21}_{-0.10}$ & $-99.00 < -99.00 < -99.00$ & -99.00 & sn \\
\enddata
\tablenotetext{a}{Passive galaxies are represented by -99.00}
\tablenotetext{b}{Redshift used in host galaxy template fitting based on SN spectra (sp), galaxy spectra (gal) or light-curve (lc)}
\end{deluxetable}

\bibliographystyle{apj}

\end{document}